\documentclass[aps,prx,twocolumn,final,letterpaper]{revtex4}

\usepackage{graphicx}   
\usepackage{import}                         % Graphics package
\usepackage{epstopdf}
\usepackage{amsmath} 
\usepackage{bm}
\usepackage{amssymb}
\usepackage{quotes}
\usepackage{transparent}
\usepackage{dcolumn}% Align table columns on decimal point
\usepackage{multirow}
\usepackage{cancel} % for the command \cancel and \cancelto
\usepackage{mdframed}
\usepackage{color}
\usepackage{bm}
\usepackage{dsfont}
\usepackage{slashed}
\usepackage{braket}
\begin{document}
\rmfamily

\title{Nonperturbative electromagnetic nonlinearities, $n$-photon reflectors, and Fock-state lasers based on deep-strong coupling of light and matter}
\author{Nicholas Rivera$^{1,2,\dagger}$, Jamison Sloan$^{3,\dagger}$, Ido Kaminer$^{4}$, and Marin Solja\v{c}i\'{c}$^{2,3}$}
\email{nrivera@fas.harvard.edu.}

\affiliation{$^{1}$Department of Physics, Harvard University, Cambridge, MA 02138, USA.  \\
$^{2}$Department of Physics, MIT, Cambridge, MA 02139, USA.  \\
$^{3}$Research Laboratory of Electronics, MIT, Cambridge, MA 02139, USA. \\
$^{4}$Department of Electrical and Computer Engineering, Technion, Haifa 32000, Israel. \\
$\dagger$ Denotes equal contribution}

\newpage

\begin{abstract}
Light and matter can now interact in a regime where their coupling is stronger than their bare energies. This deep-strong coupling (DSC) regime of quantum electrodynamics promises to challenge many conventional assumptions about the physics of light and matter. Here, we show how light and matter interactions in this regime give rise to electromagnetic nonlinearities dramatically different from those of naturally existing materials. Excitations in the DSC regime act as photons with a linear energy spectrum up to a critical excitation number, after which, the system suddenly becomes strongly anharmonic, thus acting as an effective intensity-dependent nonlinearity of an extremely high order. We show that this behavior allows for \emph{N-photon blockade} (with $N \gg 1$), enabling qualitatively new kinds of quantum light sources. For example, this nonlinearity forms the basis for a new type of gain medium, which when integrated into a laser or maser, produces large Fock states (rather than coherent states). Such Fock states could in principle have photon numbers orders of magnitude larger than any realized previously, and would be protected from dissipation by a new type of equilibrium between nonlinear gain and linear loss. We discuss paths to experimental realization of the effects described here.%, and show that some of the new nonlinearities described here could be mimicked in optics, serving as a basis for new types of lasers and other optoelectronic devices. 

%Macroscopic quantum states of light remain an important, yet demanding goal in quantum science. For example, large-number Fock states of great interest as they could enable many applications in spectroscopy, metrology, information processing, and communications. Yet, Fock states with more than even ten photons are difficult to generate. With this in mind, we present a new concept for deterministically generating mesoscopic and eventually macroscopic quantum states based on energy transfer between a gain medium and a quantum oscillator which suddenly becomes anharmonic at high energy. We consider in detail the example in which this oscillator is realized by non-perturbatively coupling a quantum emitter to an electromagnetic resonance. Beyond a threshold pumping level, this system undergoes a lasing transition. Unlike "conventional lasers", which produce approximately de-phased coherent states of the electromagnetic field, this laser produces negative-temperature photon states, which approach Fock states at larger pump strengths. The states produced by this ``Fock laser" can have excitation numbers in the hundreds while having fluctuations far below shot-noise. By virtue of the laser being in equilibrium between emission, pumping, and damping, these states do not suffer from decoherence coming from either the decay of the gain medium, or cavity losses. 
\end{abstract}
\maketitle

Recent successes in the coupling of matter and light now make it possible to realize regimes of light-matter interactions in which the coupling between light and matter can be much stronger than in established optical technologies. Because of the central role the physics of light and matter plays in many fields, these new coupling regimes are being intensely explored. One such example is the ultra-strong coupling regime, where the coupling energy is within an order of magnitude of the bare energies of the light and matter \cite{kockum2019ultrastrong}. Such regimes promise to give rise to new chemical processes \cite{hutchison2012modifying, flick2018strong, ruggenthaler2018quantum, thomas2019tilting}, strong modifications of transport and thermodynamic properties of materials \cite{orgiu2015conductivity, paravicini2019magneto}, new phases of matter, quantum simulators, and quantum technologies more broadly \cite{kockum2019ultrastrong, forn2019ultrastrong}.

Taking these ideas to their logical extreme is the so-called \emph{deep-strong coupling regime} (DSC), where the strength of the coupling is \emph{greater} than the bare energies of the light and matter. In the past few years, the first experiments in this regime have emerged \cite{yoshihara2017superconducting, forn2017ultrastrong}. Much of the interest in ultra-strong and deep-strong coupling is focused on the properties of the ground state of either one or many emitters coupled to a cavity mode, leading to many interesting new phenomena such as light-matter decoupling \cite{de2014light, rivera2019variational}, population collapses and revivals \cite{casanova2010deep}, large Lamb shifts leading to inversion of qubit energy levels \cite{yoshihara2017superconducting, yoshihara2018inversion}, and renormalization of qubit energy levels by a photonic continuum \cite{forn2017ultrastrong}. Likely, many of the potential applications of this regime have yet to be identified.

Here, we consider the opportunities afforded to us by the excited states of a DSC system, which are important from the perspective of quantum and nonlinear optics. For example, the emission of light in such systems probes the excited states. First, we show that deep-strong coupling of a two-level system to a resonant cavity leads to the formation of excitations (``photonic quasiparticles'' \cite{rivera2020light}, which we refer to as ``DSC photons'') with nonlinear properties much different than those in any known system. Then, we analyze the coupling of an emitter to this nonlinear photonic quasiparticle. We find that the coupling of an excited two-level system to this nonlinear system enables a phenomenon of $N$-photon blockade in which $N$ excitations can be populated, but $N+1$ cannot. We show that a laser or maser based on stimulated emission of DSC photons behaves fundamentally differently from a conventional maser or laser. Specifically, this maser produces close approximations to Fock states in its steady-state, rather than coherent states, as in conventional lasers. They could have a few hundred photons, thus being orders of magnitude larger than any Fock states realized thus far. Moreover, Fock states produced by this mechanism are stable against dissipation as they arise from a new type of equilibrium between nonlinear gain and linear loss. Our results may thus help to address the long-standing problem in quantum science of generating Fock states. Finally, we discuss how the concept developed here can be implemented in superconducting qubit platforms.%, and moreover how the techniques presented here can inspire new kinds of lasers with unusual quantum fluctuation properties, even at optical frequencies (without deep strong coupling).

%Perhaps the most exciting results arise when the light-matter coupling is much stronger than the bare energies of the system. While this regime is yet to be realized, recent work pushes in this direction \cite{yoshihara2018inversion}. Moreover, given the recent trend in increasing the coupling \cite{forn2019ultrastrong}, as well as the general consensus that the coupling could eventually exceed $g = 10$ $-$ if not 20 \cite{devoret2007circuit} -- we will consider these couplings. 

%Let us now consider this system in the so-called deep strong coupling (DSC) regime, which is defined such that $g \gtrsim \omega$ \cite{kockum2019ultrastrong, forn2019ultrastrong}.  This ``exotic'' regime was realized experimentally in recent experiments $-$ both based on the coupling of superconducting flux qubits to either a microwave resonator \cite{yoshihara2017superconducting} or waveguide \cite{forn2017ultrastrong}. We will be particularly interested in the regime for which $g/\omega \gg 1$. 

 %In the main text, we stick to fairly simple equations, and retain an intuitive and pedagogical style. The main technical developments are 

\section{Nonlinear photonic quasiparticles based on deep strong light-matter coupling}

\begin{figure}[t]
    \centering
    \includegraphics[width=0.5\textwidth]{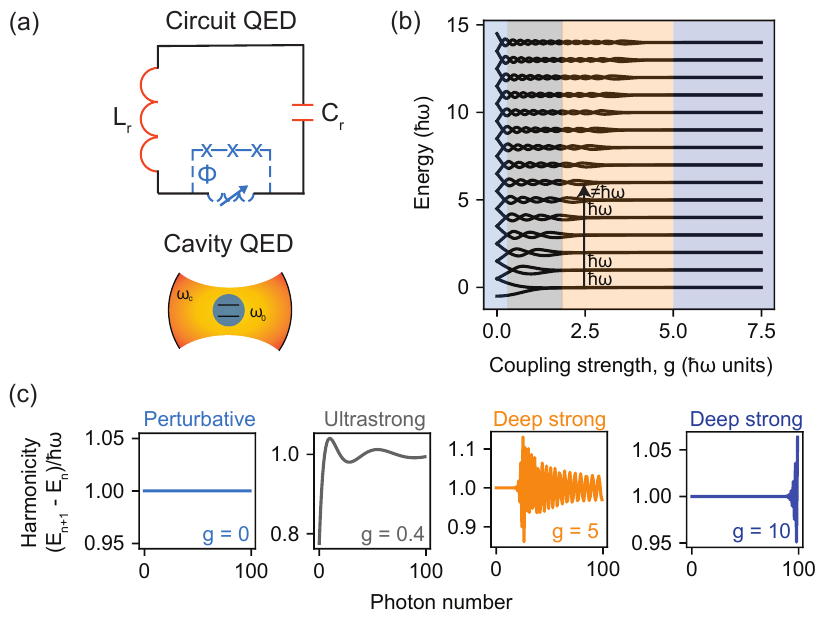}
    \caption{\textbf{High-order nonlinearities in deep strong coupling of light and matter.} (a) Schematic of a two-level system coupled to a single resonator mode, as in circuit or cavity QED. (b) Spectrum of the system from weak ($g = \tilde{g}/\omega \ll 1$) to deep-strong coupling ($g \gg 1$). Here, $\lambda = 0$. (c) Successive excitation energies for a single spin sector for different coupling values. For $g \gg 1$ the excitation energies are constant, as for a bare photon. At large photon number, they deviate rapidly and nonlinearly from harmonicity, akin to a photon with a strongly intensity-dependent nonlinearity. }
    \label{fig:dsc1}
\end{figure}

Fundamental to our results is the spectrum of a two-level system (qubit) interacting with a single-mode cavity (schematically illustrated in Fig. 1a), which we review here \cite{irish2007generalized}. The Hamiltonian, referred to as the (generalized) Rabi Hamiltonian, is given by 
\begin{equation}
    H_{\text{Rabi}}/\hbar = \frac{1}{2}\left(\omega_0\sigma_z + \lambda \sigma_x \right) + \omega a^{\dagger}a + \tilde{g}\sigma_x(a+a^{\dagger}).
\end{equation}
Here, $\omega_0$ is the transition frequency of the two-level system, $\sigma_{x,z}$ are the $x$ and $z$ Pauli matrices, $\omega$ is the cavity frequency, $a^{(\dagger)}$ is the cavity annhilation (creation) operator, and $\tilde{g}$ is the Rabi frequency. It will be convenient to non-dimensionalize the coupling as $g = \tilde{g}/\omega$. We have also generalized the standard Rabi Hamiltonian by including a term $\lambda\sigma_x$ which is relevant in contexts of superconducting qubits with applied bias fluxes \cite{yoshihara2017superconducting}. For simplicity of presentation, we consider the case of $\lambda = 0$, which leads to approximately degenerate spin states (and in which case the qubit frequency is $\omega_0$). In the SI, and in various numerical results, we do consider the effect of a finite $\lambda$ term, which yields the same qualitative conclusions. 

While the Rabi Hamiltonian cannot be analytically diagonalized in general, an approximate spectrum can be found for the regime $g \gg 1$, which forms the basis for our analytical theory. In the SI, it is shown that the approximate eigenstates are labeled by an oscillator quantum number $n = 0,1,2,\dots$ and a spin quantum number $\sigma = \pm 1$. These eigenstates $|n\sigma\rangle$ and corresponding energies $E_{n\sigma}$, for $g \gg 1$, are given by
\begin{align}
    |n\sigma\rangle &= \frac{1}{\sqrt{2}}\left(D^{\dagger}(g)|n,x+\rangle + \sigma  D^{\dagger}(-g)|n,x-\rangle \right) \nonumber \\
    E_{n\sigma}/\hbar &= \omega(n + \frac{\sigma}{2}e^{-2g^2}L_n(4g^2)),
\end{align}
where $D(z) \equiv \exp\left[z( a^{\dagger} - a) \right]$ is the displacement operator, and $L_n$ is the Laguerre polynomial of order $n$. The state $|n\rangle$ on the right-hand side refers to the Fock basis of the cavity, while the states $| x\pm\rangle$ refer to the $x$-polarized spin states of the qubit. The spectrum is plotted in Fig. 1b (adding an $g$-dependent offset $\hbar\omega g^2$ for convenience). As seen in Eq. (2), the spectrum in the DSC regime is organized into two oscillator-like ladders (one for each spin). Moreover, for large $g$, the spectrum appears almost completely harmonic, indicating the existence of an effective photon (or photonic quasiparticle, which we will sometimes call a DSC photon). To understand this, we note that for $g \gg 1$, the $\sigma_z$ acts as a perturbation to the remaining Hamiltonian, $H_{\text{DSC}}/\hbar \equiv  \omega a^{\dagger}a + g\sigma_x(a+a^{\dagger}) = \omega (b^{\dagger}b - g^2)$, where $b = D^{\dagger}(g\sigma_x)aD(g\sigma_x) = a + g\sigma_x $. This approximate Hamiltonian admits a harmonic spectrum, in which the new oscillator variables $b$ obey canonical commutation relations $[b,b^{\dagger}] = 1$, and excitations are constructed by applying further $b^{\dagger}$ operators. In other words, the eigenstates of $H_{\text{DSC}}$ are Fock states of $b$, or equivalently, displaced Fock states of $a$.

The $\sigma_z$ term breaks the even spacing of the ladder, leading to an anharmonicity (equivalently, nonlinearity) which we now quantify. Without loss of generality, we will focus on the lower-energy $\sigma=-1$ ladder, enabling us to omit the spin index in our notation. We assess the ``harmonicity'' of the spectrum by plotting successive excitation energies $E_{n+1}-E_n$ as a function of $n$, as in Fig. 1c (in units of $\hbar\omega$). We will refer to $n$ as the ``photon number.'' For strong and ultrastrong coupling, the spectrum is anharmonic at the level of a single photon, leading to the familiar phenomenon of photon blockade. For deep-strong coupling, the behavior is quite different: the spectrum is harmonic up to some critical excitation number ($n_c \sim g^2$), and then rapidly becomes anharmonic. This may be seen directly from the properties of $L_n(x)$.

To understand the relation of this strong anharmonicity to existing nonlinear optical systems, recall that a single-mode cavity, with a Kerr nonlinear medium inside of it, can be described by a Hamiltonian of the form $H_{\text{Kerr}}= \hbar\omega\left(a^{\dagger}a + \beta a^{\dagger 2} a^2\right)$ \cite{drummond1980quantum, walls2007quantum}, with $\beta$ a (typically small) dimensionless coefficient which is proportional to the refractive index shift induced by a single photon. In such a system, the energy to add an excitation is $E_{n+1}-E_n = \hbar\omega(1+2\beta n)$, meaning that the deviation from harmonic behavior is linear in the intensity (proportional to photon number). A cavity with this third order nonlinearity would have its resonance frequency shift with intensity. Thus, the plots of Fig. 1c, for a photon in a Kerr medium, would be straight lines with slope $2\beta$. This linear dependence arises from a low-order expansion of the nonlinear medium polarization in the cavity electric field: in the case of Kerr, the leading order nonlinear term would be third order, leading to a refractive index dependent on intensity. In contrast, for the case of DSC, the excitation energies do not vary linearly in photon number, but instead are very high-order near the critical photon number (as the function $L_n(4g^2)$ is practically exponential for $n \sim g^2$): as if the cavity contained a nonlinear medium whose polarization had a non-perturbative dependence on intensity. 

In what follows, we show how this non-perturbative nonlinearity arising from the spectrum of the Rabi model leads directly to the possibility of new effects, such as (1) systems which become reflective to light when a large but fixed number of excitations are in the cavity ($N$-photon blockade), and (2) lasers which generate intracavity Fock states in their steady states, as opposed to coherent states, as conventional lasers do.

%In this regime, we may write $H_{\text{Rabi}} = H_{\text{DSC}} + V$, with $V/\hbar = \frac{\omega_0}{2}\sigma_z$. In other words, the qubit energy behaves as a perturbation to the exactly diagonalizable $H_{\text{DSC}}$. 

%putting stuff here for now since I think we decided that it made conceptual sense to put this before the laser stuff
\section{Coherent driving and $N$-photon blockade in a system with deep-strong coupling}

The nonlinearity perspective presented here, although not previously noted in the literature, is largely based on the known spectrum of DSC systems. We now use this perspective to develop the main new results of this paper. 

First, we establish how these systems become reflective to light when a large, but fixed number of photons are present in the cavity. In other words, we demonstrate the fundamental phenomenon of $N$-photon blockade. To do so, we illustrate the dynamics of DSC photons under coherent driving by an external signal (e.g., an applied microwave signal, or an external laser, at frequency $\omega$). Thus, to the Hamiltonian of Eq. (1), we add a driving term of the form $H_{\text{drive}} = \eta (X^{(+)} e^{-i\omega_p t} + X^{(-)} e^{i\omega_p t})$, where $\eta$ and $\omega_p$ are the strength and frequency of the drive. Additionally, $X^{(\pm)}$ are the positive and negative frequency components of the operator $X = b + b^\dagger$, where $b$ is the annihilation operator of the DSC boson defined earlier. Specifically, the positive frequency component is defined in terms of $X$ as $X^{(+)} = \sum_{m<n} X_{mn} \ket{m}\bra{n}.$ The negative frequency operator is then $X^{(-)} = (X^{(+)})^\dagger.$ 

\begin{figure}
    \centering
    \includegraphics[scale=0.9]{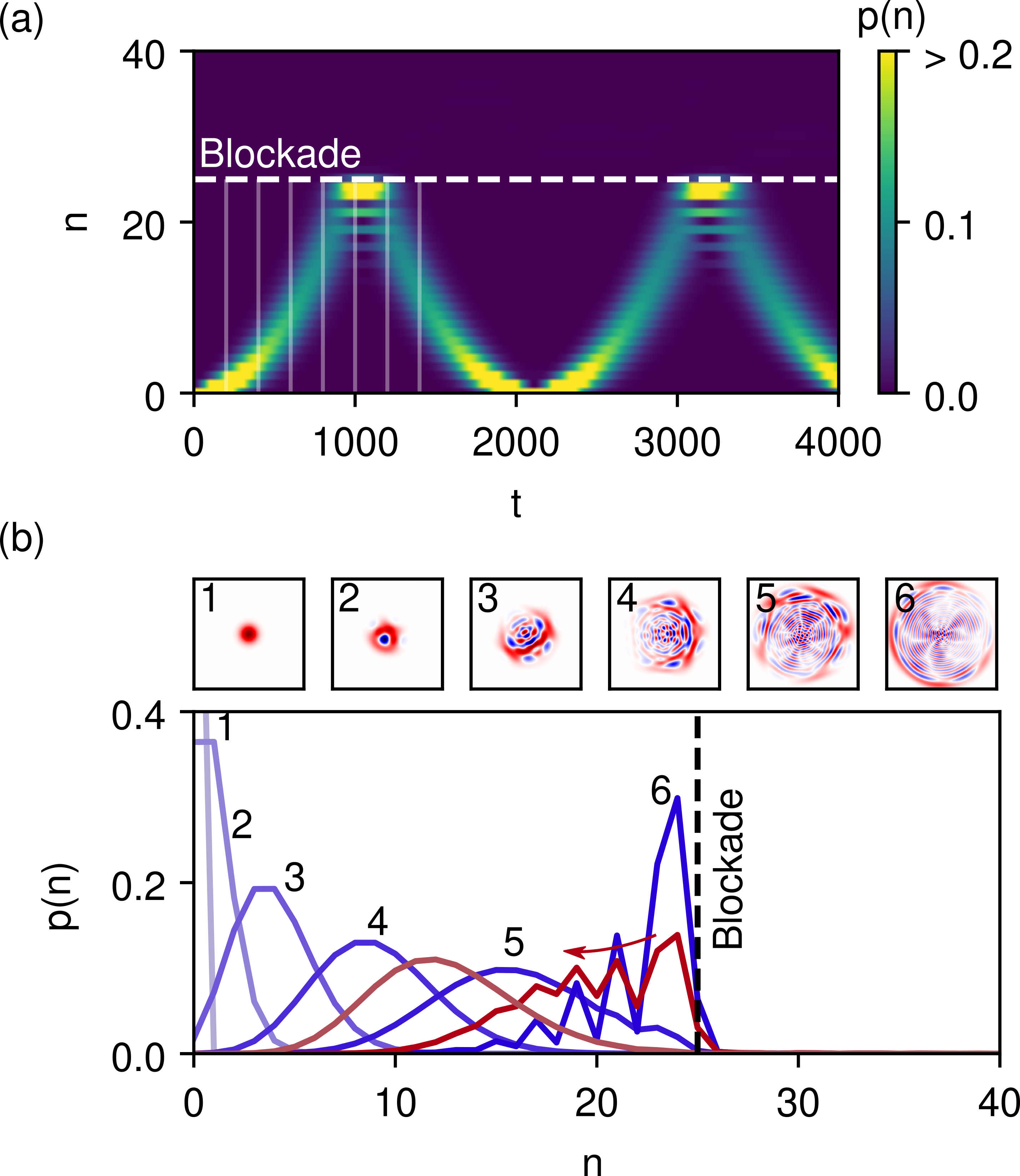}
    \caption{\textbf{Coherent pumping of DSC bosons leads to $N$-photon blockade.} (a) Probability distribution $p(n)$ of DSC photon number $n$ as a function of time $t$ in the presence of a coherent driving field. The mean photon number initially grows in accordance with a harmonic spectrum, but is abruptly stopped at the blockade number $N$ due to the sudden anharmonicity in the energy spectrum. (b) Probability distribution slices and Wigner functions at selected times (shown by vertical lines in panel (a)). The probability distribution initially evolves as an approximate coherent state, but then acquires a reduced variance at the blockade point. Interference fringes in $p(n)$ appear due to the nonlinear squeezing that occurs. After the blockade point, the distribution turns around as it is reflected due to the blockade. Parameters used are $g=5$, $\lambda = 0.1$, $\eta = 0.005$, and $\omega_p$ set to the difference between the two lowest energy eigenvalues in the ``down'' manifold of spin states.}
    \label{fig:dsc_coherent}
\end{figure}

In Fig. \ref{fig:dsc_coherent}a, we show the probability distribution $p(n)$ for the number of DSC photons $n$, as a function of the time $t$ after the coherent drive is turned on. We note that $n$ enumerates over states in the ``down'' manifold of spin states. For the parameters we use, the population which leaks out into the ther manifold is small. Fig. \ref{fig:dsc_coherent}b shows slices of this probability distribution, and the corresponding Wigner functions for the DSC photon (for $\sigma = -1)$. Immediately after the pump is turned on, the average excitation number begins to grow in a manner which is similar to that of a ``normal" (i.e. linear) pumped cavity. This is expected, since for low excitation numbers, the DSC bosons have an almost perfectly harmonic spectrum. However, as soon as the probability distribution approaches the strongly anharmonic point $N \sim g^2$, the probability distribution begins to compress, as it becomes harder for photons to be added to the cavity. The driving frequency which was once resonant for lower photon numbers becomes highly non-resonant at the blockade point, resisting the population of excitations into any state beyond $N$. As the blockade point is approached, the quantum state of light deviates further from the classical coherent state which is produced with a linear resonance, as evidenced by the Wigner functions which take on a negative (blue) value with many fringes. 

Once the blockade point is hit, the distribution actually \emph{turns around}, and then repeats the cycle. This extreme form of the behavior occurs when the timescale associated with the field growth is faster than those associated with dissipation in the system. In the SI, we show the influence of dissipation. When this dissipation is present, the dynamics are similar for short times to what is shown in Fig. 2: after dissipation begins to act, the system reaches a steady state which can be squeezed in DSC excitation number, having a variance which is below the mean (steady-state squeezing tends to be fairly modest in this configuration, about 3 dB).

\section{Light emission in the deep-strong coupling regime}
\begin{figure*}[t]
    \centering
    \includegraphics[width=\textwidth]{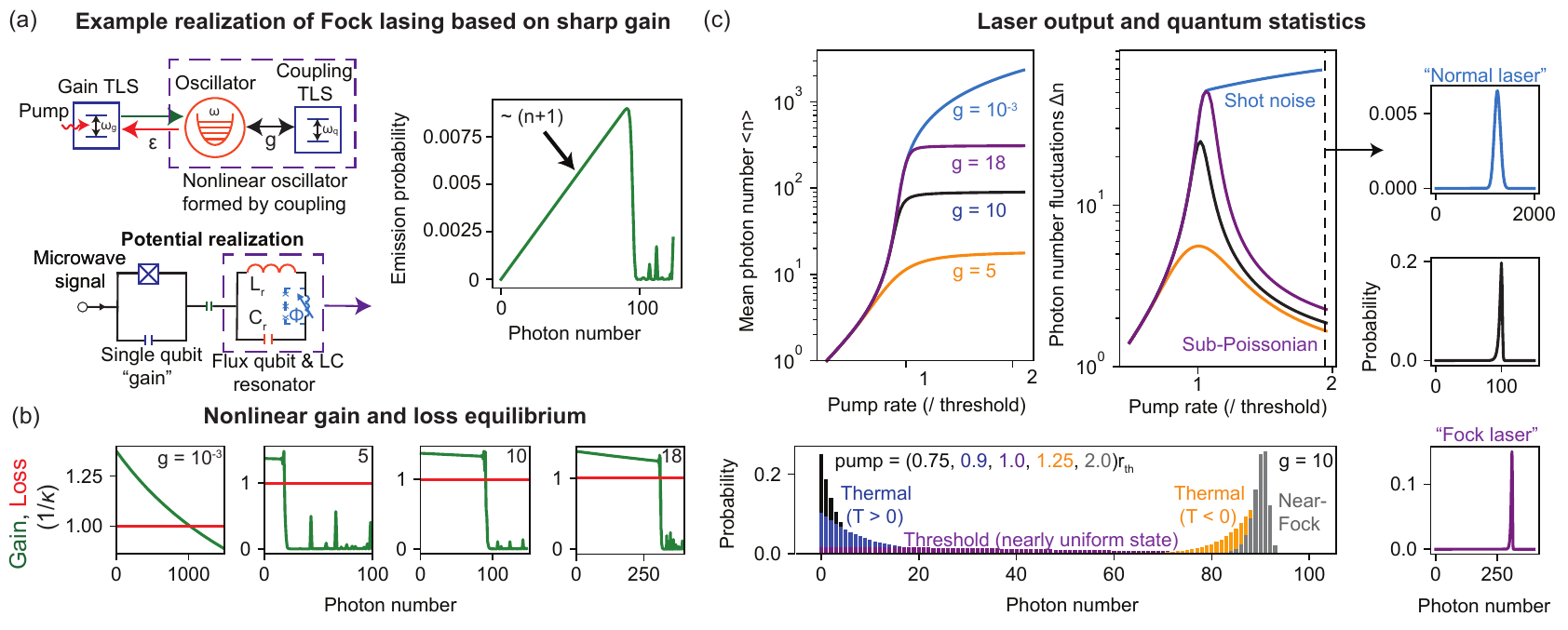}
    \caption{\textbf{Fock lasing due to equilibrium between high-order nonlinearity and dissipation.} (a) Light emission of DSC photons can be understood in terms of the coupling of an emitter (e.g., a probe qubit) weakly coupled to the DSC system, as might be realized by coupling a superconducting qubit to a flux-qubit-LC-resonator system. The probability to stimulatedly emit DSC photons scales as $n+1$ for small $n$, and then sharply decreases due to the sudden anharmonicity for $n > n_c \sim g^2$. ``TLS'' denotes two-level system. (b) This behavior leads to a gain medium whose gain coefficient (green lines) is highly nonlinear. The quantum state of DSC photons will depend on how this nonlinear gain comes into equilibrium with the loss (red lines).  (c) Steady-state intensity and power fluctuations of lasers in different coupling regimes as a function of pump intensity. For the “harmonic” regimes (weak, and deep-strong), a rapid growth in intensity at threshold is seen. In contrast to the weak coupling regime (as in a “normal” laser; light blue curve), a laser operating in the deep-strong coupling regime has its intensity saturate, and its fluctuations vanish at high pump, converging to a high-number Fock state (dark blue and purple curves), leading to Fock-like statistics (right). (c, bottom) Statistics for different pump strengths for a single coupling, showing evolution from thermality to Fock-like statistics.}
    \label{fig:dsc1}
\end{figure*}
%We will show that a light emitter coupled to this highly nonlinear ``DSC photon'' undergoes a qualitatively different type of stimulated emission that, when integrated with feedback, lead to light sources such as masers that produce Fock and other unusual states. 
Specifically, we study how light emission is modified by these photonic quasiparticles. Unlike most studies of light emission with photonic quasiparticles (reviewed for example in \cite{rivera2020light}), here we look at the unique modifications coming from the nonlinear properties. Consider an external qubit (denoted `em', for emitter) coupled to this DSC photon. The exact form of the coupling depends on the circuit implementation. To keep the discussion concrete, we will consider a simple coupling Hamiltonian of the type
\begin{align}
 H &= H_{\text{Rabi}} + \frac{\omega^{\text{em}}_0}{2}\sigma^{\text{em}}_z + V \nonumber \\
 V/\hbar &= \epsilon  \sigma^{\text{em}}_x (b + b^{\dagger}) \approx \epsilon (\sigma^{\text{em}}_+ b + b^{\dagger}\sigma^{\text{em}}_-),
\end{align}
which couples the emitter directly to the DSC photon. Regarding the assumed form of the Hamiltonian, we note that our conclusions are not particularly sensitive to the exact form of the interaction \footnote{The term $ \sigma^{\text{em}}_x (b + b^{\dagger})$ contains an interaction between the dipole moment of the emitter and that of the qubit. Such interactions are to be generically expected, as especially emphasized in recent works on superradiant phase transitions, as well as gauge invariance in ultrastrong coupling cavity and circuit quantum electrodynamics \cite{jaako2016ultrastrong, de2018breakdown, de2018cavity, di2019resolution, settineri2019gauge}. We could write the term in question as $\alpha  \sigma^{\text{em}}_x \sigma_x. $ Here, $\alpha = 2\epsilon g$. Because this dipole-dipole term leads only to changes in spin quantum number, and not changes in excitation number (see SI), and because the spins are not separated by $\omega$, these terms have little effect on the dynamics of the photon number probabilities that we consider. For example, we find that ignoring this term altogether leads to the same conclusions. Hence, for the purposes of the manuscript, we have taken a simple coupling that illustrates the physics best (emission of a ``b'' particle by an emitter).}. What we do assume however is that $\epsilon$ is small, so that the coupling of the external emitter to the DSC system is weak ($\epsilon \ll \omega$). Thus, the system in mind is a single resonator coupled to two qubits, one with weak coupling and one with deep strong coupling, as illustrated in Fig. 2a. 

To understand emission and absorption of DSC photons, consider the case in which the qubit is in its excited state $|e\rangle$ and there are $n$ DSC photons present of spin $-1$ (e.g., occupying the state $|n,-1\rangle$ of Eq. (2)). If the qubit is at frequency $\omega$ (same as in Eq. (2)), then the qubit transition will be nearly resonant with the transition $n \rightarrow n+1$ of the DSC photon, provided $n \lesssim n_c$. The dynamics can be restricted to the subspace $\{|e,n\rangle, |g,n+1\rangle \}$, and the probability of (stimulated) emission $P(n+1)$ is simply given by 
\begin{align}
P(n+1) &= \frac{(n+1)\epsilon^2}{\Delta_{n+1}^2+(n+1)\epsilon^2}\sin^2\left(\sqrt{\Delta_{n+1}^2+(n+1)\epsilon^2}t \right) \nonumber \\
\Delta_{n+1} &= \frac{\omega}{2}\left(\delta - \frac{1}{2}e^{-2g^2}(L_n(4g^2) - L_{n+1}(4g^2))\right).
\end{align}
Here, $\delta$ is the dimensionless detuning of the emitter and $\omega$, such that $\omega^{\text{em}}_0 - \omega \equiv \omega\delta$. Eq. (4) is the direct consequence of the Jaynes-Cummings dynamics of a two-level system (emitter) with a boson (DSC photon) with some detuning. The detuning depends on excitation number due to the nonlinearity of the DSC photon, and the detuning sharply rises near $n_c$ (Fig. 1c). In Fig. 2a, we plot the stimulated emission probability as a function of $n$ after a small evolution time $t \ll \epsilon^{-1}$ and for $\delta = 0$. For $n < n_c$, $\Delta_{n+1} \approx 0$, that probability is simply $(n+1)(\epsilon t)^2$, corresponding to stimulated emission proportional to $n+1$, as expected for conventional photons. For $n \gtrsim n_c$, the emission probability drops rapidly, because of the corresponding rapid increase in $\Delta_n$. This can be understood as a type of $N$-photon blockade, in which a system can readily accept $N$ excitations, but not $N+1$. For $N=1$, this corresponds to the conventional photon blockade observed and discussed extensively in strong coupling cavity QED (in which the polariton anharmonicity is strong at the level of one photon). In the ``conventional'' strong-coupling cavity QED case, this $N=1$ blockade corresponds to single-photon nonlinearity which is highly desired for many applications. Here, the spectrum is such that the photon blockade is \emph{delayed} to $N = N_c$ photons, leading to an exotic and strong quantum nonlinearity that operates at $N \gg 1$ photons.

\subsection{A new type of laser}
\begin{figure}[t]
    \centering
    \includegraphics[width=0.48\textwidth]{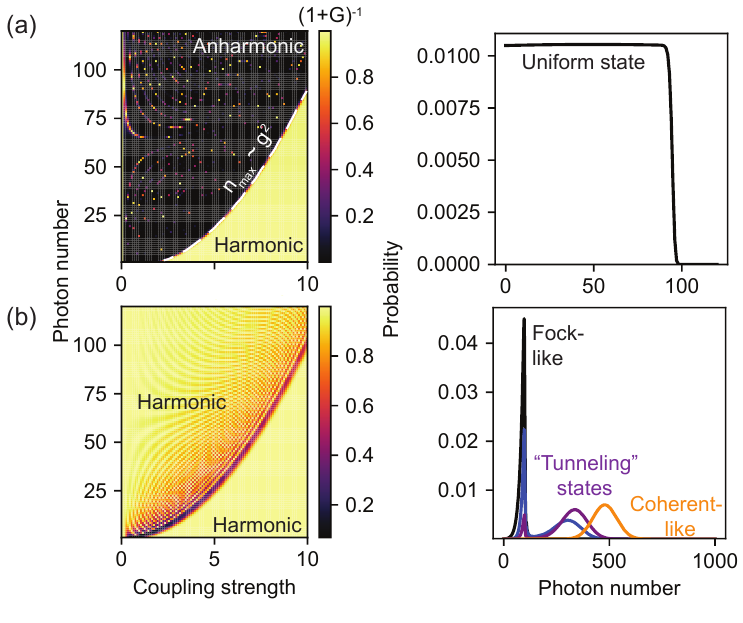}
    \caption{\textbf{Uniform and tunneling states arising at the harmonic-to-anharmonic crossover.} (a) The function $(1+G_n))^{-1}$ which propagates the photon probability distribution from $n$ to $n+1$, plotted as a function of coupling strength and photon number ($G_n = F_n/\Gamma$). A harmonic-to-anharmonic crossover occurs for a maximum photon number $n_{\text{max}} \sim g^2$ for which the propagation function goes to zero. When this happens, the probability of having photons larger than $n_{\text{max}}$ vanishes. Near threshold (where the effective temperature of the photon goes to infinity), this leads to nearly uniform states of the electromagnetic field sharply cutoff at the maximum photon number (right panel).  (b) When the decay rate of the gain medium is large, the anharmonic region becomes narrower (bottom left), and for sufficient pump intensity, the photon distribution can ``tunnel'' through the barrier, evolving effectively as a coherent state. In this tunneling regime, the distribution becomes bimodal, taking on the characteristics of the Fock and coherent states for some pump parameters (bottom right). }
    \label{fig:fig4}
\end{figure}
Eq. (4) displays one of the main results: that the high-order nonlinearities arising from non-perturbative quantum electrodynamical coupling lead to a type of gain (stimulated emission) that is correspondingly non-perturbative in intensity. One may imagine that this type of nonlinear stimulated emission would have implications for lasers -- or in this case, masers, given that the most imminent implementations, based on circuit QED, would be at microwave frequencies. We will stick to the term ``laser'' since it has largely subsumed masers. In this section, we show that the DSC-based gain discussed before creates lasing into
high-order Fock states (rather than coherent states). 

We now show how the nonlinear gain provided by the coupling of an excited two-level system to DSC photons can result in a laser with new steady state photon statistics. To capture the resulting lasing dynamics in a quantum mechanical way, we shall find an equation of motion for the reduced density matrix $\rho$ of the DSC photon (tracing out the gain medium). This equation takes into account both the stimulated emission dynamics and the loss dynamics associated with, for example, leakage from the cavity (which we take here for simplicity as the primary loss mechanism for the DSC photon). In the SI, we derive the equation using several methods, all in agreement with each other. Here, we focus on the equation for the DSC photon occupation probabilities, $\rho_{nn}$. Assuming that excited states of the gain medium are pumped at rate $r$, the equation of motion for the DSC photon density matrix is found to be:
\begin{align}
    \dot{\rho}_{nn} &= R_n n \rho_{n-1,n-1}  - \left(R_{n+1}(n+1)\rho_{nn} + \kappa_n n \rho_{nn}\right) \nonumber \\ &+ \kappa_{n+1} (n+1) \rho_{n+1,n+1} 
\end{align}
Here, $R_n = \frac{2r\epsilon^2}{\Gamma^2 + F_n}$ is the stimulated emission coefficient, with $F_n \equiv  4n\epsilon^2 + \frac{1}{4}\omega^2 e^{-4g^2}(L_{n}(4g^2) - L_{n-1}(4g^2))^2$. The $R_n$ are plotted (green curves) in Fig. 2b for different coupling strengths. For weak coupling, it is simply saturable gain $R(n) \sim 1/(1+n/n_s)$ with $n_s$ the saturation photon number. For DSC, we see that the gain coefficient is given by the standard saturable form for $n < n_c$ and then rapidly decays for $n \gtrsim n_c$ (with occasional oscillations arising from the oscillatory behavior of the Laguerre polynomials). Here, $\kappa_n = \kappa|\langle n-1|a+a^{\dagger}|n\rangle|^2$, with $\kappa$ the decay rate of the cavity in the absence of DSC (see SI for derivation). We note that for simplicity, the gain medium has been taken to have population and coherence decay rates arising from the same source (so that $\Gamma = 1/T_1 = 2/T_2$). This simplifies the calculations but does not qualitatively change our conclusions.

The steady state photon probability distribution is entirely different from that of a traditional laser, which produces a dephased coherent state. To quantify this, we solve a recursion relation to obtain the steady-state probability distribution $\rho_{n,n} = Z^{-1}\prod\limits_{m=1}^n R_m/\kappa_m$, with $Z$ a normalization coefficient enforcing $\sum_n \rho_{n,n} = 1$. In Fig. 2c, we show the intra-cavity photon number and photon fluctuations for DSC in comparison with weak coupling. We also present the corresponding photon statistics. In the weak coupling regime, the photon number as a function of pump follows the canonical “S-curve” relating the input pump and output intensity of a laser. The output intensity grows sharply for pump beyond the threshold pump level, $r_{\text{th}} = \kappa\Gamma^2/2\epsilon^2$. The fluctuations below threshold are essentially those of a thermal state, and far above threshold, grow according to shot-noise (as  $\sqrt{n}$, as for a Poissonian distribution corresponding to a randomly-phased coherent state): this is the textbook result of the laser theory of Lamb and Scully \cite{scully1967quantum,scully1999quantum}. In contrast, the “Fock laser” ($g = 5, 10, 18$), saturates (at $n_c \sim g^2$), and the photon number fluctuations go to zero, leading to the quantum statistics of a Fock state (Fig. 2c, right) as the pump increases. Fig. 2c (bottom) further shows how the photon statistics evolve with pump and coupling (taken for $g=10$; additional results shown in SI). Beyond threshold, the distribution of photons (for DSC) approaches that of a thermal state of negative temperature. Such states, as the pump is increased (and $T \rightarrow 0^-$), approach states where only the highest-most level is filled, with minimal spread, which closely approximates a Fock state of $n_c$ DSC photons.

To understand this Fock lasing effect, it is helpful to refer to the gain and loss curves of Fig. 2b, as well as the steady-state distribution $\rho_{n,n} = Z^{-1}\prod\limits_{m=1}^n R_m/\kappa_m$. The steady-state distribution has the property that the probabilities are maximized where gain equals loss, and probabilities are suppressed if one of gain or loss far exceeds the other. In particular, the larger the angle between the gain and loss curves (at the crossing point), the tighter the concentration of probabilities about the mean. Increasing the pump rate $r$ further will scale the gain curve up, leading to a steeper slope and further suppression of photon number fluctuations, leading asymptotically to a Fock state.

Beyond these close approximations to high-photon-number Fock states, other unusual states can arise from the equilibrium between gain and loss, due to this sudden anharmonicity for $n \gtrsim n_c$. For example, near threshold, where the number fluctuations increase dramatically, the resulting distribution is nearly step-like, going to zero rapidly for $n_c$. This anharmonicity provides a ``wall'' for the photon probability distribution that is too hard to pass through, even as the fluctuations get very large near threshold. These effects also depend on the decay rate of the gain medium: if the decay rate is high, then it provides gain over a large bandwidth, and so changes in the DSC photon frequency have a reduced effect on the stimulated emission rate $R_n$. As a result, for increasing pump, the distribution can ``tunnel'' through the wall, leading to states that interpolate between Fock and coherent states, as well as pure coherent states for large enough pump.

% We should emphasize that while the application of this laser theory used in a much different context may not necessarily seem justified, these results can be found through other means. One method, which we show in the SI, is direct numerical evaluation of the steady-state of the Liouvillian, which shows the same features as the analytical solutions presented here.
\section{Discussion and outlook}

\textbf{Comparison to other approaches} The new type of infinite-order nonlinearity realized by the Rabi Hamiltonian in the deep-strong coupling regime, in principle, enables the deterministic creation of large Fock states, which has proven challenging in general. In this section, we briefly review other approaches to generating Fock states, especially those relevant at microwave frequencies. It is currently possible to deterministically produce Fock states of order roughly 15 in microwave resonators through a combination of external driving of the cavity by microwave pulses and superconducting transmon qubits \cite{heeres2017implementing}. Fock states have also been generated in microwave cavities by strongly coupling them to transmon qubits that are repeatedly pumped to inject photons into the cavity at deterministic times \cite{hofheinz2008generation}. Older foundational work in the field of cavity quantum electrodynamics made use of Rydberg atoms strongly coupled to microwave cavities in order to generate low-order Fock states using principles such as the one above, as well as quantum feedback protocols \cite{ rempe1990observation, varcoe2000preparing, sayrin2011real}. Such Rydberg atom-cavity interactions form the basis for new theoretical proposals to extend microwave Fock states to higher photon numbers \cite{uria2020deterministic, canela2020bright}. 

Compared to these approaches, the approached outlined here has some notable advantages. The ``Fock laser'' shown here, which populates a cavity with $N$ excitations based on stimulated emission, does so because of a strong anharmonicity of the spectrum. As a result, any approximately monochromatic pump of energy which incoherently pumps the cavity can lead to population of a Fock state. This is in contrast to approaches such as the ``micromaser'' in which Rydberg atoms, which  interact with the cavity in the conventional strong-coupling regime, are injected into a cavity one-at-a-time. There, Fock states can in principle based on the concept of ``trap states'' \cite{rempe1990observation, varcoe2000preparing}: for a certain photon number $n$ and interaction time between atom-and-cavity $\tau$, the probability of adding a photon can vanish (due to a vanishing matrix element), leading to a fixed point where a Fock state of order $n$ can be populated. This effect therefore is not robust against loss, multiple atoms being present at a time, inhomogeneities in the interaction time, etc. This is why it has proven difficult to produce good approximations to high-number Fock states in practice.

Compared to transient approaches such as in \cite{hofheinz2008generation,heeres2017implementing} approximate $N$-photon state is a steady-state solution, even in the presence of loss. As a result, the Fock state can be maintained for arbitrarily long, in contrast to ``transient'' approaches, in which the Fock state lasts only for the cavity lifetime. 

\textbf{Implementation:} Of course, the primary limitation in our approach has to do with using deep-strong light-matter coupling, which requires extremely strong light-matter coupling strengths. That said, recent work on realizing deep-strong coupling of superconducting qubits to a microwave (LC) resonator, as in \cite{yoshihara2017superconducting, yoshihara2018inversion}, provides a path to observing the effect predicted here. It is already possible to have control over $g$ from weak coupling to a value of nearly 2. With a $g$ of 2, one can see from Fig. 1 that a Fock state of three or four excitations could be pumped. For smaller $g$, in the ultra-strong coupling regime where $0.1 < g < 1$, only one excitation can be created, as a manifestation of the conventional photon blockade effect \cite{ridolfo2012photon, le2016fate}. Thus, the behavior of our model from weak to (modest) deep strong coupling can already be realized. 

Regarding the gain medium, it is important to point out that while a typically gain medium consisting of many emitters, the physics can also be realized by a gain medium consisting of a single qubit. The qubit should be weakly coupled to the same cavity as the strongly coupled qubit, and will lase, provided that the gain from this one qubit is above threshold \cite{scully1999quantum}. Single-qubit gain is responsible for much of the exciting experiments on ``one-atom lasers'' (in real \cite{mckeever2003experimental, an1994microlaser} and artificial atoms \cite{liu2015semiconductor, stehlik2016double, burkard2020superconductor}), in which a single atom or artificial atom provides enough gain to lase. 

Thus, a conceptually simpler $-$ and perhaps more attractive $-$ approach to realize our predictions is to consider a gain medium consisting of a continuously pumped superconducting qubit which is \emph{weakly} coupled to the same resonance as the strongly coupled qubit (which for example happens if $\epsilon \ll \kappa$). In Fig. 2, we took $\epsilon = 10^{-5}\omega$, and $\Gamma = 10^{-3}\omega$. Thus, for a single gain qubit, threshold is reached provided the quality factor of the resonator is above $5 \times 10^6$. There are two advances that would support reaching larger $g$ values: the rapidly increasing coupling constants that have been realized with superconducting qubits (see Fig. 1 of \cite{forn2019ultrastrong}), and early estimates in this field suggesting the possibility of $g$ values of roughly 20 \cite{devoret2007circuit}. Another important point is that while we have focused in this paper on incoherent pumping (based on emission from two-level systems), the nonlinear emission physics described in this manuscript could also be extended to coherent pumping of the DSC photon by an external microwave signal. In that case, we expect that by combining the high-order nonlinearity of the DSC photon with a frequency-dependent leakage loss (e.g., loss coming from a reflection filter), one could engineer a highly nonlinear loss which would be ``dual'' to the highly nonlinear gain introduced in Fig. 2.

Summarizing, we have shown a physical principle – using non-perturbative photonic nonlinearity – which could enable lasers that produce deterministic, macroscopic quantum states of light, such as Fock states. Part of the new physics uncovered here, related to lasing in systems with sharply nonlinear gain, could in principle also be extended into the optical regime. In fact, in \cite{pontula2022strong,rivera2023creating} $-$ inspired by the developments in this manuscript $-$ we discuss how trying to mimic the new ``Fock lasers'' predicted here, but at \emph{optical frequencies}. This is done essentially by combining a highly frequency-dependent loss with Kerr nonlinearities to get an effectively non-perturbatively nonlinear loss.  Thus, the principles established here, independently of deep-strong coupling, should also give rise to new ideas and experiments in the optical domain. %much of the new physics here, arising from the non-perturbative dependence of gain on intensity, can in fact inspire new kinds of devices at optical frequencies (not using deep-strong coupling).

\section{Acknowledgements}

N.R. acknowledges the support of a Junior Fellowship from the Harvard Society of Fellows, as well as earlier support from a Computational Science Graduate Fellowship of the Department of Energy (DE-FG02-97ER25308), and a Dean’s Fellowship from the MIT School of Science. J.S. was supported in part by the Department of Defense NDSEG fellowship no. F-1730184536. This material is based upon work supported in part by the Air Force Office of Scientific Research under the award number FA9550-20-1-0115; the work is also supported in part by the US Army Research Office through the Institute for Soldier Nanotechnologies at MIT, under Collaborative Agreement Number W911NF-18-2-0048.

\bibliographystyle{unsrt}
\bibliography{Fock_laser.bib}

% These results provide a fundamentally new scheme to produce mesoscopic and eventually macroscopic quantum states of the electromagnetic field, and can be extended to "lasers" based on other quantum fields. 

% Our findings reveal how lasers based on non-perturbative (deep-strong) light-matter interaction behave fundamentally differently from conventional lasers, where the light-matter interaction is weak. 

\end{document}

\rmfamily

\title{Supplemental Materials: \\ Nonperturbative photonic nonlinearity and Fock-state lasers based on \\ deep-strong coupling of light and matter}
\author{Nicholas Rivera$^{1}$, Jamison Sloan$^{2}$,  Ido Kaminer$^{3}$, and Marin Solja\v{c}i\'{c}$^{1,2}$}

\affiliation{$^{1}$ Department of Physics, MIT, Cambridge, MA 02139, USA. \\
$^{2}$ Research Laboratory of Electronics, MIT, Cambridge, MA 02139, USA. \\
$^{3}$ Department of Electrical Engineering, Technion, Haifa 32000, Israel.}

\maketitle

\noindent	

\noindent
\tableofcontents

\clearpage 

\section{Fock lasing based on deep-strong light-matter coupling}

In this Supplement, we derive and extend the results of the main text. Consider a system involving matter coupled to a cavity mode very strongly, so that the system is in the ultra- or deep-strong coupling regime. This system is described by the Rabi Hamiltonian of Eq. (1) of the main text (Hamiltonian and variables re-defined here for self-containedness):
\begin{equation}
    H_{\text{Rabi}}/\hbar = \frac{1}{2}\left(\omega_0\sigma_z + \lambda \sigma_x \right) + \omega a^{\dagger}a + \tilde{g}\sigma_x(a+a^{\dagger}),
\end{equation}
Here, $\omega_0$ is the transition frequency of the two-level system, $\sigma_{x,z}$ are the $x$ and $z$ Pauli matrices, $\omega$ is the cavity frequency, $a^{(\dagger)}$ is the cavity annhilation (creation) operator, and $\tilde{g}$ is the Rabi frequency. We also non-dimensionalize the coupling as $g = \tilde{g}/\omega$. 

Let us now transfer energy into this system by means of external emitters, treated as two-level systems of energy $\omega_0$. Let us assume the emitter is primarily interacting with the cavity (as it is too far for direct interactions with the dipole of the matter). Let us then take the full Hamiltonian describing the coupling of one emitter to the light-matter system as
\begin{equation}
   H/\hbar =  \frac{\omega^{\text{em}}_0}{2}\sigma^{\text{em}}_z +  H_{\text{Rabi}} + \epsilon\sigma_{x,\text{em}}(b+b^{\dagger}),
\end{equation}
which couples the emitter directly to the DSC photon. We can also consider interactions solely between the emitter and the resonator field, replacing $b \rightarrow a$. We consider this case as well, to show that the exact nature of the emitter-qubit dipole-dipole coupling does not qualitatively change our conclusions.

If the emitter is in the excited eigenstate $\ket{e}$, and it is resonant with a transition of the Rabi Hamiltonian, the emitter can transfer energy to the light-matter system. Upon interaction with a second emitter, if the next transition of the Rabi model has nearly the same frequency, the system can get further excited. A key observation is that in the deep-strong coupling regime $g \gg \omega$, the eigenstates are approximately equally spaced, and the excitations are oscillator-like, quite similarly to the zero-coupling case. This should allow the possibility of reaching a very high excitation number in the presence of many emitters, based on stimulated emission of these oscillator modes (We will call them DSC photons). When the coupling is not infinite, as in a realistic case, the levels are no-longer fully equally spaced. This detuning is photon-number dependent, thus acting as a nonlinearity which may qualitatively change the steady-state of this type of laser.  

To begin, we need to derive simple forms for the eigenstates of the Rabi Hamiltonian in the deep-strong coupling limit. Then, we will consider their coupling to external emitters, and write a coarse-gained equation of motion for the density matrix of the DSC bosons, and then solve it. 

\subsection{Eigenstates of the Rabi Hamiltonian}

In what follows, we will take $\omega_0 = \omega$ (resonant) and $\lambda = 0$. In later subsections, we will analytically and numerically consider the case of a finite $\lambda$, which is found to preserve our main findings.

In the deep-strong coupling regime, we can treat the matter term in the Rabi Hamiltonian as a perturbation to the remainder of the Hamiltonian. The remainder of the Hamiltonian (divided by $\hbar$), which we call $H_{\text{DSC}}$ is
\begin{equation}
    H_{\text{DSC}} = \omega a^{\dagger} a + \Tilde{g}\sigma_x(a+a^{\dagger}) = \omega(a^{\dagger} + g\sigma_x)(a + g\sigma_x) - \omega g^2,
\end{equation}
where $g \equiv \Tilde{g}/\omega$ is a dimensionless measure of the coupling strength. Introducing the displacement operator $D(g\sigma_x) = \exp\left[g\sigma_x ( a^{\dagger} - a) \right]$, where we've taken $g$ real without loss of generality, we have
\begin{equation}
    H_{\text{DSC}} = \omega D^{\dagger}(g\sigma_x)a^{\dagger} a D(g\sigma_x),
\end{equation}
where we've omitted the overall constant $-\omega g^2$. From here, we can easily see that the eigenstates of this Hamiltonian are of the form $D^{\dagger}(\pm g)\ket{\pm x, n}$, where $\ket{x}$ denotes the x-spin basis, and $n$ is a Fock state. In other words, the eigenstates involve the spin being \emph{x}-polarized (rather than \emph{z}-polarization), and the photon being in a \emph{displaced} Fock state (rather than just a Fock state). Clearly,
\begin{equation}
    H_{\text{DSC}}D^{\dagger}(\pm g)\ket{\pm x, n} = \omega D^{\dagger}(\pm g)a^{\dagger} a D(\pm g)D^{\dagger}(\pm g)\ket{\pm x, n} = n\omega D^{\dagger}(\pm g)\ket{\pm x, n}.
\end{equation}
Clearly then, in this limit, the eigenstates are evenly spaced, and doubly degenerate. In fact, it can be seen as a system of two non-interacting bosons (``DSC photons''). Introducing $b_{\sigma} = a + g\sigma_x$ we can write the Hamiltonian as 
\begin{equation}
    H_{\text{DSC}} = \omega b^{\dagger} b.
\end{equation}
It can also be easily seen that $[b,b^{\dagger}] = 1$. 

The degeneracy of the DSC Hamiltonian is split by the matter Hamiltonian. We can find the resulting eigenstates and eigenenergies using degenerate first-order perturbation theory. The ``good'' eigenbasis of the problem is
\begin{equation}
    \ket{n,\sigma} = \frac{1}{\sqrt{2}}\left(D^{\dagger}\ket{+x,n} + \sigma D \ket{-x,n} \right),
\end{equation}
where $\sigma = \pm 1$, and a displacement operator without an argument implies that the argument is $g$. The energies of the resulting states are
\begin{align}
    E_{n\sigma} &= \frac{\omega}{2}\bra{n,\sigma}\sigma_z\ket{n,\sigma} \nonumber \\
    &= \frac{\omega}{4}\left(\bra{+x,n}D + \sigma \bra{-x,n}D^{\dagger} \right)\sigma_z \left(D^{\dagger}\ket{+x,n} + \sigma D \ket{-x,n} \right) \nonumber \\ 
    &= \sigma\frac{\omega}{4}\left(\bra{n}D^2\ket{n} + \bra{n}D^{\dagger 2}\ket{n}\right) \nonumber \\ &= \sigma\frac{\omega}{2}\bra{n}D^2\ket{n} 
    \equiv \sigma\frac{\omega}{2}D_n,
\end{align}
where $D_n = \bra{n}D^2(g)\ket{n} = \bra{n}D(2g)\ket{n}$. These eigenstates and energies are sufficiently accurate, even for $g=2$ or $g=3$. 

\subsubsection{Evaluation of $D_n$. }

Let us evaluate the $D_n$. To do so, we write:
\begin{align}
    D_n(2z) &= \bra{n} D D \ket{n} = \frac{1}{n!}\bra{0} a^nD Da^{\dagger n} \ket{0} = \frac{1}{n!}\bra{0} D(D^{\dagger}a^nD)(Da^{\dagger n}D^{\dagger})D \ket{0} \nonumber  \\ &= \frac{1}{n!}\bra{-z} (a + z )^n(a^{\dagger}-z^*)^n \ket{z}.
\end{align}
To proceed, insert a ``complete'' set of states using the over-completeness of the coherent states. That leaves us with 
\begin{equation}
    D_n(2z) = \frac{1}{\pi n!}\int d^2\alpha \bra{-z} (a + z )^n\ket{\alpha}\bra{\alpha}(a^{\dagger}-z^*)^n \ket{z} = \frac{1}{\pi n!}\int d^2\alpha  (\alpha + z )^n(\alpha^*-z^*)^n \langle \alpha | z\rangle  \langle -z|\alpha \rangle.
\end{equation}
Using the rule for the overlap of two coherent states, we have 
\begin{equation}
    D_n(2z) = \frac{1}{\pi n!}\int d\alpha d\alpha^*  (\alpha + z )^n(\alpha^*-z^*)^n e^{-\alpha\alpha^* - zz^* + \alpha^* z - z^* \alpha},
\end{equation}
where we have written things this way to emphasize that $\alpha$ and $\alpha^*$ are independent variables. We can now write this as 
\begin{equation}
    D_n(2z) = \frac{e^{zz^*}}{\pi n!}\int d\alpha d\alpha^*  (\alpha + z )^n(\alpha^*-z^*)^n e^{-\alpha\alpha^* + z(\alpha^*- z^*) - z^*( \alpha + z)}.
\end{equation}
Then, we transform variables as $\alpha \rightarrow \alpha - z$ and $\alpha^* \rightarrow \alpha^* + z^*$ to get 
\begin{equation}
    D_n(2z) = \frac{e^{zz^*}}{\pi n!}\int d\alpha d\alpha^* ~ \alpha^n\alpha^{*n} e^{-(\alpha-z)(\alpha^* + z^*) + z\alpha^* - z^*\alpha} = \frac{e^{2zz^*}}{\pi n!}\int d\alpha d\alpha^* ~ \alpha^n\alpha^{*n} e^{-\alpha\alpha^*  + 2z\alpha^* - 2z^*\alpha}.
\end{equation}
This can be generated from simpler integrals by differentiation, as:
\begin{equation}
    D_n(2z) = (-1)^n\frac{e^{2zz^*}}{\pi n!}\frac{\partial^{2n}}{\partial^n(2z)\partial^n(2z^*)}\int d\alpha d\alpha^* ~ e^{-\alpha\alpha^*   + 2z\alpha^* - 2z^*\alpha}.
\end{equation}
Completing the square in the remaining integral gives
\begin{align}
    D_n(2z) &= (-1)^n\frac{e^{2zz^*}}{\pi n!}\frac{\partial^{2n}}{\partial^n(2z)\partial^n(2z^*)}e^{-4zz^*}\int d\alpha d\alpha^* ~ e^{-\alpha\alpha^*   + 2z\alpha^* - 2z^*\alpha + 4zz^*} \nonumber \\ &= (-1)^n\frac{e^{2zz^*}}{\pi n!}\frac{\partial^{2n}}{\partial^n(2z)\partial^n(2z^*)}e^{-4zz^*}\int d\alpha d\alpha^* ~ e^{-(\alpha - 2z)(\alpha^* + 2z^*)}.
\end{align}
Shifting variables as $\alpha \rightarrow \alpha + 2z$ and $\alpha \rightarrow \alpha - 2z^*$, and performing the final Gaussian integral, we have 
\begin{equation}
     D_n = (-1)^n\frac{e^{2zz^*}}{ n!}\frac{\partial^{2n}}{\partial(2z)^n\partial(2z^*)^n}e^{-4zz^*}.
\end{equation}
These are related to Laguerre polynomials. To see this, take the derivative with respect to $z^*$. We will also use the notation $z \rightarrow x/2$ and $z^* \rightarrow y/2$ for clarity.
\begin{equation}
     D_n(2z) = (-1)^n\frac{e^{xy/2}}{ n!}\frac{\partial^{2n}}{\partial x^n \partial y^n}e^{-xy} = \frac{e^{xy/2}}{n!}\frac{\partial^{n}}{\partial (xy)^n}(xy)^n e^{-(xy)} = e^{-xy/2} \left(\frac{e^{xy}}{n!}\frac{\partial^{n}}{\partial (xy)^n}(xy)^n e^{-(xy)}\right).
\end{equation}
From the Rodrigues formula for the Laguerre polynomials, we then have
\begin{equation}
    D_n(2z) = e^{-2|z|^2}L_n(4|z|^2).
\end{equation}
Given that the $z=g$, we have then that the the level splitting is given by
\begin{equation}
    E_{n\sigma} = n\omega + \sigma\frac{\omega}{2}e^{-2|g|^2}L_n(4|g|^2)
\end{equation}

\subsection{Time-evolution of the coupled system}

With the approximate eigenstates of the Rabi Hamiltonian, we now want to understand the full dynamics of $H$. We will take advantage of the fact that for a laser, $\epsilon$ is small, and in particular, $\epsilon \ll \omega$, so that the rotating-wave approximation is valid. In this system, the rotating wave approximation consists of only considering the dynamics within degenerate subspaces of the unperturbed Hamiltonian 
\begin{equation}
    H_0 = \frac{\omega_0}{2}\sigma_{z,\text{em}} +  H_{\text{Rabi}}.
\end{equation}
The eigenstates of the problem are $\ket{k}\ket{n\sigma}$, where now $ k = 0,1$ denotes emitter states (ground is zero, excited is one). The energies of such states are (up to a shift)
\begin{equation}
    E_{kn\sigma} = (n+k(1+\delta))\omega + \sigma\frac{\omega}{2}D_n.
\end{equation}
where we have taken $\omega_0 = (1+\delta)\omega$. From here on out, let us assume  $\delta \ll \omega$. In that case, it is easy to see that the following four states form our nearly degenerate subspace:
\begin{equation}
   \{ \ket{1,n-1,+}, \ket{0,n,+}, \ket{1,n-1,-}, \ket{0,n,-} \}.
\end{equation}
We now need to understand the action of the interaction Hamiltonian $V \equiv \epsilon \sigma_{x,\text{em}}(b+b^{\dagger})$ on this subspace. First of all,
\begin{equation}
    \bra{k'n'\sigma'}V\ket{kn\sigma} = 0 ~\text{if}~ k = k'.
\end{equation}
For $k = -k'$, we have 
\begin{align}
    \bra{-kn'\sigma'}V\ket{kn\sigma} &= \frac{\epsilon}{2}\left(\bra{+x,n'}D + \sigma' \bra{-x,n'}D^{\dagger} \right)D^{\dagger}(g\sigma_x)(a+a^{\dagger})D(g\sigma_x) \left(D^{\dagger}\ket{+x,n} + \sigma D \ket{-x,n} \right) \nonumber \\ 
 &= \epsilon\frac{1+\sigma\sigma'}{2}(\sqrt{n}\delta_{n',n-1}+\sqrt{n+1}\delta_{n',n+1}).
\end{align}
From these matrix elements, we see that: if the pseudo-spin ($\sigma$) is conserved, then a non-zero matrix element occurs only when the boson number changes by 1. Noting that $b = a + g\sigma_x$, we can also readily describe interactions using $a$ as $\langle n'\sigma' | a + a^{\dagger}|n\sigma\rangle = \sqrt{n}\delta_{n',n-1}+\sqrt{n+1}\delta_{n',n+1} - 2g\frac{1-\sigma\sigma'}{2}\delta_{nn'}$, such that: when the pseudo-spin changes, non-zero matrix elements occur only when the boson number is conserved. When the spin is conserved, the matrix elements are the same as for $b+b^{\dagger}$. Since only states with different photon number differ appreciably in frequency (and in particular, will be resonant with the emitter we introduce), the interactions are effectively the same whether we describe $a$ or $b$. This is also to say that any modification in the coefficient of the dipole-dipole interaction between emitter and qubit will lead to the same result insofar as DSC photon dynamics are concerned. In Fig. S1, we show the matrix elements of $a$ and $b$ between adjacent states of the same spin, as well as $a^{\dagger}a$ and $b^{\dagger}b$. For $n < n_c \sim g^2$, they behave as one might expect for an oscillator.

We should note that beyond $n_c$, these states and matrix elements that we calculate based on degenerate perturbation theory are expected to change significantly. However, the approximate result turns out to describe the system well because the probabilities to find photon numbers beyond $n_c$ are strongly suppressed in the Fock laser, rendering the description relatively insensitive to these details.

\begin{figure}[h]
    \centering
    \includegraphics[width=0.9\textwidth]{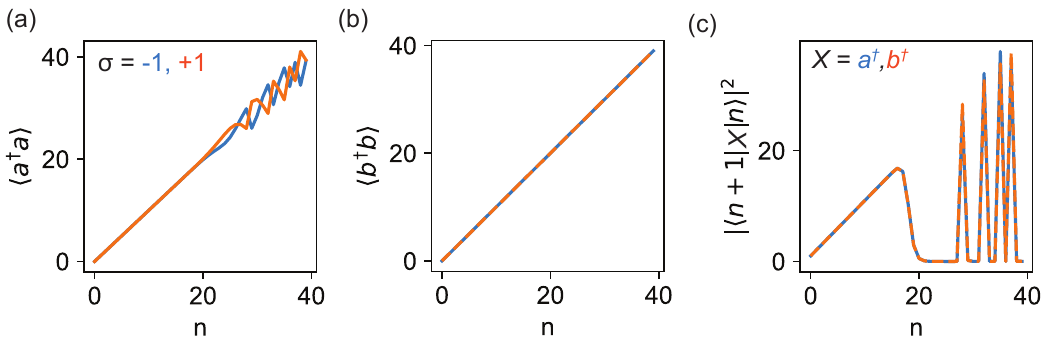}
    \caption{Matrix elements of $a^{\dagger}a$, $b^{\dagger}b$ and $a, b$, showing that $b$ counts excitations of the DSC system over the full range of eigenstates. However, up to $n_c$, $a$ and $b$ act similarly.}
\end{figure}

Based on these considerations, we see that the Hamiltonian in the degenerate subspace may be written as:
\begin{align}
   H &= (\delta\omega + \frac{\omega}{2}D_{n-1})\ket{1,n-1,+}\bra{1,n-1,+} + ( \frac{\omega}{2}D_{n})\ket{0,n,+}\bra{1,n,+} \nonumber \\
   &+ \epsilon\sqrt{n}\ket{1,n-1,+}\bra{0,n,+} + \text{h.c} \nonumber \\ 
   &+ (\delta\omega - \frac{\omega}{2}D_{n-1})\ket{1,n-1,-}\bra{1,n-1,-} +  (-\frac{\omega}{2}D_{n})\ket{0,n,-}\bra{0,n,-} \nonumber \\
   &+ \epsilon\sqrt{n}\ket{1,n-1,-}\bra{0,n,-} + \text{h.c}.
\end{align}
As we can see here, there are two independent blocks of the Hamiltonian (for each pseudo-spin) and we can thus study them separately. Let us assume that we're at zero temperature, and so the ground state has the $-$ pseudo-spin, which we assume to be conserved for all times. In that case, we can work with the simple 2x2 Hamiltonian
\begin{equation}
    H_{\text{eff}} = \omega(\delta - \frac{1}{2}D_{n-1})\ket{1,n-1}\bra{1,n-1}  - \frac{1}{2}\omega D_{n}\ket{0,n}\bra{0,n} + \epsilon\sqrt{n}\ket{1,n-1}\bra{0,n} + ~\text{h.c},
\end{equation}
where the pseudo-spin label has been dropped. This can be written in terms of Pauli matrices as
\begin{equation}
    H_{n} = \frac{\omega}{2}(\delta - \frac{1}{2}(D_{n-1}+D_{n}))I + \frac{\omega}{2}(\delta - \frac{1}{2}(D_{n-1} - D_{n}))\sigma_z +\epsilon\sqrt{n}\sigma_x.
\end{equation}
Introducing $E_{n} = \frac{\omega}{2}(\delta - \frac{1}{2}(D_{n-1}+D_{n}))$, $\Delta_n = \frac{\omega}{2}(\delta - \frac{1}{2}(D_{n-1} - D_{n}))$, we have very simply
\begin{equation}
    H_{n} = E_{n}I + \Delta_n \sigma_z +\epsilon\sqrt{n}\sigma_x.
\end{equation}
As we will see in the next section, we need to know how states of the form $\ket{1,n-1}$ evolve over time. Thus we need
\begin{equation}
    e^{-it(\Delta_n \sigma_z +\epsilon\sqrt{n}\sigma_x)} \equiv e^{-i|U_n|t(\hat{U}_n\cdot\sigma)},
\end{equation}
with $\hat{U}_n = \frac{(\Delta_n,0,\epsilon\sqrt{n})}{\sqrt{\Delta_n^2+n\epsilon^2}}$ and $|U_n| = \sqrt{\Delta_n^2+n\epsilon^2}$. Since $\hat{U}\cdot\sigma = 1$, we have 
\begin{equation}
e^{-i|U_n|t(\hat{U}_n\cdot\sigma)} = \cos(|U_n|t) -i(\hat{U}_n\cdot\sigma)\sin(|U_n|t).
\end{equation}
Therefore
\begin{equation}
    e^{-iHt/\hbar}\ket{1,n-1} = (\cos(|U_n|t)-i\hat{U}_{nz} \sin(|U_n|t))\ket{1,n-1} - i\hat{U}_{nx} \sin(|U_n|t)\ket{0,n},
\end{equation}
So, the probability of remaining in the same state is
\begin{equation}
    P(1,n-1) = \cos^2(|U_n|t) + \frac{\Delta_n^2}{\Delta_n^2+n\epsilon^2} \sin^2(|U_n|t)  = 1 - \frac{n\epsilon^2}{\Delta_n^2+n\epsilon^2}\sin^2(|U_n|t),
\end{equation}
while the probability of transitioning is
\begin{equation}
    P(0,n) = \frac{n\epsilon^2}{\Delta_n^2+n\epsilon^2}\sin^2(|U_n|t).
\end{equation}

\section{Equation of motion for DSC photons}

Now we consider the description of laser action. To do so, we formulate an equation for how the density matrix of the DSC photon changes due to stimulated emission by the emitter. The method of analysis presented closely follows the coarse-grained density matrix technique used to describe conventional lasers. It is described in many books, such as \cite{scully1999quantum,walls2007quantum}. Suppose we have our emitter coupled to the light-matter (DSC) system. The emitter unit starts in the state $|i\rangle$ and the DSC system is taken to have a density matrix $\rho_{\text{DSC}}$, so that the initial density matrix of the total system $\rho_{\text{tot}}$ is given by $\rho_{\text{tot}}(t) = |i\rangle\langle i|\rho_{\text{DSC}}(t)$. Let us look for an equation describing only the evolution of the DSC system. Assuming the interaction over a time $T$ corresponds to the evolution operator $U$, we have that
\begin{equation}
    \rho_{\text{tot}}(t+T) = U(T)|i\rangle\langle i|\rho_{\text{DSC}}(t) U^{\dagger}(T).
\end{equation}
Let us express all operators in terms of their matrix elements, writing the above equation as 
\begin{equation}
    \rho_{\text{tot}}(t+T) = \sum\limits_{ff',mm',nn'} \rho_{\text{DSC},nn'}(t)\langle fm |U(T)|in\rangle\langle in'| U^{\dagger}(T)|fm'\rangle |fm\rangle\langle f'm'|,
\end{equation}
where $\rho_{\text{DSC},nn'}$ denotes the matrix elements of the DSC system, for a fixed pseudo-spin. More compactly,
\begin{equation}
    \rho_{\text{tot}}(t+T) = \sum\limits_{ff',mm',nn'} \rho_{\text{DSC},nn'}(t) U_{fm,in}(T) U^*_{f'm', in'}(T) |fm\rangle\langle f'm'|,
\end{equation}
The DSC photon density matrix, $\rho_{\text{DSC}} = \text{tr}_{\text{em}} \rho_{\text{tot}}$, can then be expressed as
\begin{equation}
    \rho_{\text{DSC}}(t+T) = \sum\limits_k \langle k| \rho_{\text{DSC}}(t+T) |k\rangle =  \sum\limits_{k, mm',nn'} \rho_{\text{DSC},nn'}(t) U_{km,in}(T) U^*_{km', in'}(T) |m\rangle\langle m'|.
\end{equation}

From here, a number of approaches can be followed. If there is no loss in the system, then the density matrix of the total system upon the next iteration is simply  $\rho_{\text{DSC}}(t+T) = |i\rangle\langle i|\rho_{\text{DSC}}(t+T)$ and this procedure can be iterated in a discrete fashion. The evolution can also be seen as continuous if, over time $T$, the change in the density matrix is small. This doesn't describe the early stages of the evolution, but it can describe later stages once there are many bosons in the system. If there is a steady state, then the continuous evolution must describe the run-up to the steady state, as changes get smaller over time. In such a case, we have
\begin{equation}
    \dot{\rho}_{mm'} = r\left(\sum\limits_{k,nn'} U_{km,in}(T) U^*_{km', in'}(T) - \delta_{n,m}\delta_{n',m'} \right)\rho_{nn'},
\end{equation}
where $r = N/T$ is the number of excited emitters introduced into the system in time $T$. We have also dropped the ``DSC'' subscript for the DSC photon for brevity. These terms in the evolution of the density matrix describe the gain in the system. In addition, since there are losses associated with the cavity, the emitter, and the matter coupled to the cavity, we need to describe those. For simplicity, we will assume the emitter has loss, and so does the DSC photon, but not the matter (qualitatively similar results arise if the matter has loss). 

\subsection{Lindblad terms}

Here we describe the effect of dissipation of the DSC photon on the equation of motion for its density matrix. Let's assume for simplicity that the cavity loss the primary source of dissipation in the problem. For weak coupling, the standard prescription is to add a Lindblad term to the Liouvillian which prescribes the evolution of the density matrix. The Lindblad term would be (at zero temperature) $\mathcal{D}[a]\rho \equiv -\frac{\kappa}{2}(a^{\dagger}a\rho + \rho a^{\dagger}a - 2a\rho a^{\dagger})$. As is well known from studies of dissipation in ultra-strong coupling of light and matter, the use of the standard Lindblad term leads to unphysical excitations (in the energy eigenbasis), even at zero temperature, and zero pumping \cite{beaudoin2011dissipation}. Part of the issue is that in the USC regime, the $a$ operator can create excitations in the eigenbasis, clearly not representing dissipation. Framed in terms of the standard derivation, the issue could be said that the interaction picture $a$ operator has negative frequencies, and the use of white noise (with frequencies $-\infty$ to $\infty$) introduces contributions from these negative frequencies \cite{ridolfo2012photon, kockum2019ultrastrong}. The issue can be rectified by keeping in mind the positive-frequency nature of the reservoir. 

We now use this procedure to describe dissipation in the deep-strong coupling regime. Although the technique has been worked out for ultra-strong coupling, there is a commonly used assumption in the final result that all transitions have different frequencies, which does not necessarily hold in DSC, when the energy ladder is quasi-harmonic. Interestingly, as we will show from a physical dissipator, the issues described above create much less error in the DSC regime, and the use of an operator like $a$ or $b$ produces a similar result to a proper positive-frequency jump operator, as their negative frequency parts get exponentially suppressed. 

Let us consider the Lindblad term arising from a system-bath coupling of the form
\begin{equation}
    V = J\sum\limits_k (V_kb_k + V^*_kb_k^{\dagger}),
\end{equation}
where $J$ is a DSC system operator (e.g., $a+a^{\dagger}$ or $b+b^{\dagger}$), and the $b_k$ are the bath operators, satisfying $[b_k, b_{k'}^{\dagger}] = \delta_{kk'}$. The couplings $V_k$ between system and bath are weak. To isolate the positive-frequency parts of $J$, we express it in its energy eigenbasis as $J = \sum\limits_{n>m} J_{mn}T_{mn} + \sum\limits_{m>n} J_{mn}T_{mn}  + \sum\limits_{n} J_{nn}T_{nn} \equiv J^{(+)} + J^{(-)} + J^{0} $ , with $J_{mn} = \langle m |J|n\rangle$ and $T_{mn} = |m\rangle \langle n|$. 

In what follows, we will consider the bath to be concentrated around $\omega$, but broadband enough that the white-noise approximation may be made for any transitions we consider. For example, a bath with a half-bandwidth of 10\% of $\omega$ would be sufficient for the values of $g, V_k$ we consider. It would include all active transitions of the form $n \rightarrow n+1$, but would not include higher transitions (though the matrix elements for them are small anyway), and in the presence of a $\lambda$ term, it would also not include transitions that only change spin (for $\lambda = 0$, the two spins are very nearly degenerate and so the argument should be treated with more care). Therefore, we may describe the interaction of Eq. (39) within the rotating wave approximation, instead considering 
\begin{equation}
    V \approx \sum\limits_k (V_kb_k J^{(+)} + V^*_kb_k^{\dagger} J^{(-)}).
\end{equation}
We note that it is not necessary to take the RWA at this stage, but it makes the subsequent manipulations simpler.

Thus, we may approximate the evolution of the reduced density matrix of the DSC system (in the interaction picture) to second-order in time-dependent perturbation theory, as:
\begin{equation}
    \dot{\rho}_{\text{DSC},I} = -i\text{tr}_b\left([V_I(t), \rho(0)]\right) - \int\limits_0^t dt'~ \text{tr}_b\left(\left[V_I(t),\left[V_I(t'),\rho_I(t')\right]\right]\right),
\end{equation}
where $\rho_I$ is the system-bath density matrix, $\rho_{\text{DSC},I}$ is the system density matrix, $V_I$ is the system-bath coupling in the interaction picture, and $\text{tr}_b$ denotes the partial trace with respect to the bath. For simplicity, we will consider the bath at zero temperature. Upon taking the trace with respect to the bath, the term which is linear in $V_I$ will vanish, and the equation of motion becomes
\begin{equation}
    \dot{\rho}_{\text{DSC},I} = - \int\limits_0^t dt'~ \text{tr}_b\left(V_I(t)V_I(t')\rho_I(t') + \rho_I(t')V_I(t')V_I(t) - V_I(t)\rho_I(t')V_I(t') - V_I(t')\rho_I(t')V_I(t) \right).
\end{equation}
The first term may be simplified, taking the trace with respect to the bath variables, as 
\begin{equation}
    -\int\limits_0^t dt'\int\limits_0^{\infty} d\omega~ D(\omega) |V(\omega)|^2 e^{i\omega(t'-t)} J^{(-)}_I(t)J^{(+)}_I(t')\rho_{\text{DSC}}(t'),
\end{equation}
where $D(\omega)$ is the density of bath states, and we have replaced the sum over $k$ by an integral over bath frequencies. Since an operator of the form $J^{(+)}$ is a pure de-excitation operator, no spurious excitations are introduced, and the integration limits may be extended to $-\infty$. Doing so, and making the white noise approximation, one immediately finds that the term evaluates to $-\frac{\kappa}{2} J^{(-)}_I(t)J^{(+)}_I(t)\rho_{\text{DSC}}(t)$, where $\kappa = 2\pi\rho|V|^2$. A similar manipulation for the remaining terms yields that the free dissipation dynamics of the DSC Hamiltonian are governed by 
\begin{equation}
    \dot{\rho}_{\text{DSC},I} = -\frac{\kappa}{2}\left(J_I^{(-)}J_I^{(+)}\rho_{\text{DSC}} + J_I^{(-)}J_I^{(+)}\rho_{\text{DSC}} - 2J_I^{(+)} \rho_{\text{DSC}} J_I^{(-)} \right).
\end{equation}

Let us use this to find the contribution of dissipation to the equation of motion for the populations, $\rho_{nn}$. From here on out, we will suppress the ``DSC'' subscript. We will ignore the spin degree of freedom (and restrict the dynamics to a single spin ladder). Although this is not rigorous, one expects this to capture well the dynamics of the DSC photon number as, for $\lambda = 0$, one will just expect the nearly degenerate spins to be mixed, with little change of the oscillator quantum numbers. We validate this numerically. For finite $\lambda$ the spin ladders can be split appreciably, and so they will decouple. Consider a $J$ of the form $b+b^{\dagger}$. As discussed in the main text, $b$ is a pure de-excitation operator, and $b^{\dagger}$ is a pure creation operator. Therefore, $J^{(+)} = b$. Using the fact that $\langle n'\sigma | b | n \sigma\rangle = \sqrt{n}\delta_{n',n-1}$ and $b^{\dagger}b|n\sigma\rangle = n|n\sigma\rangle$, one immediately arrives at
\begin{equation}
    \dot{\rho}_{nn} = -\kappa n \rho_{nn} + \kappa(n+1)\rho_{n+1,n+1},
\end{equation}
which is similar to the form one would expect for damping of a conventional photon. This is perhaps unsurprising in light of the fact that the DSC photon is essentially harmonic up to $n_c \sim g^2$. It is worth noting that the matrix elements derived for $a, b$ in Eq. (24) are based on first-order degenerate perturbation theory. Beyond $n_c$, these approximations do not hold up and the states and matrix elements change significantly. However, the approximate result turns out to describe the system well because the probabilities to find photon numbers beyond $n_c$ are strongly suppressed. It is also worth noting that if we chose $a$ instead of $b$ as the jump operator, when we neglect spin, the matrix elements are the same. Numerically, we find that whether we choose $a$ or $b$ as the jump operator, negligible levels of excitations are created in the ground state, and the steady-state of the Fock laser we describe is not qualitatively changed. These numerics are shown in the last section.

\subsubsection{Rate equations}

We will now obtain a closed set of equations for the diagonals of the DSC density matrix, to get the probability of different Fock state occupations of the DSC photons. Setting $m=m'$, we have 
\begin{equation}
    \dot{\rho}_{mm} = r\left(\sum\limits_{k, nn'} U_{km,in}(T) U^*_{km, in'}(T) - \delta_{n,m}\delta_{n',m} \right)\rho_{nn'}.
\end{equation}
The set of equations for the coarse grained density matrix is only closed when $U_{km,in}(T) U^*_{km, in'}(T)$ is zero unless $n=n'$. In that case, we have
\begin{equation}
    \dot{\rho}_{mm} = r\left(\sum\limits_{k, n} |U_{km,in}(T)|^2 - \delta_{n,m}\delta_{n',m} \right)\rho_{nn'} = r\left(\sum\limits_{k, n} |U_{km,in}(T)|^2 \rho_{nn} - \rho_{mm} \right).
\end{equation}
We can now note the conditions under which the equations for the populations become closed. We require $U_{km,in}(T) U^*_{km, in'}(T)$ is zero unless $n=n'$. This is equivalent to saying that a transition $in \rightarrow km$ and $in' \rightarrow km$ are not simultaneously possible. Supposing $i$ is also an eigenstate of the light-matter system, and that we are in the RWA, this statement appears to amount to energy conservation, as transitions are assumed to be only efficient if they are resonant, so that $E_i + E_n = E_k + E_m$. Therefore the condition that $U_{km,in}(T) U^*_{km, in'}(T) \neq 0$ for $n\neq n'$ requires $E_n = E_{n'}$, which, for a single oscillator, requires $n=n'$. 

In the weak coupling regime then, we have (adding in the photon losses)
\begin{equation}
    \dot{\rho}_{mm} = r\sum\limits_{k, n} |U_{km,in}(T)|^2 \rho_{nn} - r\rho_{mm} + \kappa (m+1) \rho_{m+1,m+1}  - \kappa m \rho_{mm}. 
\end{equation}

Let's now consider the case of the emitter coupled to our light-matter system. Since we inject emitters in the excited state, we have $i=1$. The state $1n$ couples only to $1n$ and $0(n+1)$. So, the sum over probabilities leaves only the scattering matrix coefficients $U_{1m,1m}$ and $U_{0m,1(m-1)}$. Therefore, the coarse-grained equation simplifies to:
\begin{equation}
    \dot{\rho}_{mm} = r(|U_{0m,1(m-1)}(T)|^2 \rho_{m-1,m-1} + |U_{1m,1m}(T)|^2 \rho_{mm}) - r\rho_{mm} + \kappa (m+1) \rho_{m+1,m+1}  - \kappa m \rho_{mm}. 
\end{equation}
We found these probability coefficients when studying the dynamics of the Hamiltonian in the degenerate subspace of fixed pseudo-spin. Plugging in the results there, we have \footnote{To check a limiting case, we set the detunings are zero. In that case, $|U_n| = \sqrt{n}\epsilon$, and we have $$\dot{\rho}_{nn} = r\sin^2(\epsilon T\sqrt{n}) \rho_{n-1,n-1} -\left( r\sin^2(\epsilon T\sqrt{n+1}) + \kappa n \right)\rho_{nn} + \kappa (n+1) \rho_{n+1,n+1}$$
This coincides exactly with the equation of motion of the so-called micromaser, which describes the interaction of injected two-level atoms interacting with a cavity (in the perturbative coupling regime $g \ll \omega$). This is quite interesting as the micromaser equations assume $g \ll 1$, while here, we are starting from the limit $g \gg 1$. Moreover, by averaging over decay times as we do in the next subsection, we will find exactly the standard Scully-Lamb master equation for a conventional laser.  What's happening here is that in the weak-coupling regime, assuming the emitter is resonant, the detunings also approximately vanish between the nearly degenerate levels. And so we get a similar equation, except that it the conventional case, it is in the photon basis, and here it is in the DSC photon basis.}

\begin{equation}
    \dot{\rho}_{nn} = \frac{rn\epsilon^2}{\Delta_n^2+n\epsilon^2}\sin^2(|U_n|T) \rho_{n-1,n-1}  -\left(\frac{r(n+1)\epsilon^2}{\Delta_{n+1}^2+(n+1)\epsilon^2}\sin^2(|U_{n+1}|T) + \kappa n \right)\rho_{nn} + \kappa (n+1) \rho_{n+1,n+1}.   
\end{equation}

To proceed, we must the emitter loss ($T_1$ and $T_2$ decay) into account. Assuming that the emitter loss manifests as exponential decay with rate $\Gamma$, the effect is to average the probability coefficients over $T$ with probability distribution $P(T) = \Gamma e^{-\Gamma T}$. Noting that
\begin{equation}
    \Gamma \int dT ~e^{-\Gamma T}\sin^2(\alpha T) = \frac{2\alpha^2}{\Gamma^2 + 4\alpha^2},
\end{equation}
we have
\begin{align}
    \dot{\rho}_{nn} &= r\frac{n\epsilon^2}{\Delta_n^2+n\epsilon^2}\frac{2U_n^2}{\Gamma^2 + 4U_n^2} \rho_{n-1,n-1} \nonumber \\ & -\left( r\frac{(n+1)\epsilon^2}{\Delta_{n+1}^2+(n+1)\epsilon^2}\frac{2U_{n+1}^2}{\Gamma^2 + 4U_{n+1}^2} + \kappa n \right)\rho_{nn} \nonumber \\ &+ \kappa (n+1) \rho_{n+1,n+1}.   
\end{align}
Noting that $U_n^2 = \Delta_n^2 + n\epsilon^2$, we have
\begin{align}
    \dot{\rho}_{nn} &= r\frac{2n\epsilon^2}{\Gamma^2 + 4(\Delta_n^2 + n\epsilon^2)} \rho_{n-1,n-1} \nonumber \\ & -\left( r\frac{2(n+1)\epsilon^2}{\Gamma^2 + 4(\Delta_{n+1}^2 + (n+1)\epsilon^2)} + \kappa n \right)\rho_{nn} \nonumber \\ &+ \kappa (n+1) \rho_{n+1,n+1}.   
\end{align}
Assuming resonance between the emitter and the light-matter system, we have finally 
\begin{equation}
    \dot{\rho}_{nn} = \frac{2rn\epsilon^2}{\Gamma^2 + F(n)} \rho_{n-1,n-1}  -\left( \frac{2r(n+1)\epsilon^2}{\Gamma^2 + F(n+1)} + \kappa n \right)\rho_{nn} + \kappa (n+1) \rho_{n+1,n+1},
\end{equation}
with the nonlinearity, $F(n)$ defined as
\begin{equation}
    F(n) = 4n\epsilon^2 + \frac{1}{4}\omega^2 e^{-4g^2}(L_{n}(4g^2) - L_{n-1}(4g^2))^2.
\end{equation}
Here, we have used $\Delta_n = \frac{\omega}{2}(\delta - \frac{e^{-2g^2}}{2}(L_{n-1}(4g^2) - L_{n}(4g^2)))$ with $\delta = 0$.

\subsection{Steady-state dynamics}
Perhaps one of the most important results is the steady-state dynamics of the system. Thus we want to solve $\dot{\rho}_{n} = 0$ with the constraint $\sum\limits_n \rho_{n} = 1$ (introducing the shorthand $\rho_{n} = \rho_{nn}$). Writing the steady-state equation as $0 = A_n \rho_{n-1}  + B_n\rho_{n} + C_n \rho_{n+1}$, we have the recursion relation: $\rho_{n+1} = -\frac{B_n\rho_n + A_n\rho_{n-1}}{C_n},$
with $\rho_0 = 1$ and $\rho_{-1} = 0$. Since any scale multiple of $\rho$ also  solves this equation, we can normalize the solution at the end to satisfy the normalization constraint. 

This equation can be simplified by noting that $B_n = -(A_{n+1}+C_{n-1})$. We thus have $A_n \rho_{n-1}  - A_{n+1}\rho_{n} - C_{n-1}\rho_{n} + C_n \rho_{n+1} = 0$ or alternatively
\begin{equation}
    A_n \rho_{n-1}  - C_{n-1}\rho_{n}  = A_{n+1}\rho_{n} - C_n \rho_{n+1}.
\end{equation}
Defining the difference $S_n = A_n \rho_{n-1}  - C_{n-1}\rho_{n}$, we see that $S_n = S_{n+1}$. Since $S_0 = A_0\rho_{-1} - C_{-1}\rho_0 = 0$, we have that $S_n = 0$ for all $n$, and thus the simpler recursion relation:
\begin{equation}
   \rho_{n+1} = \frac{A_{n+1}}{C_n}\rho_{n} \implies \rho_n = \left(\prod\limits_{m=1}^n \frac{A_m}{C_{m-1}}\right)\rho_0.
\end{equation}
The initial $\rho_0$ is taken as 1 understanding that we must normalize the probability distribution at the end of the calculation. Plugging in the forms of the $A$ and $C$ coefficients, we have 
\begin{equation}
    \rho_n = \frac{1}{Z}\prod\limits_{m=1}^n  \frac{2r\epsilon^2/\kappa}{\Gamma^2 + F(m)} \equiv \frac{1}{Z}\prod\limits_{m=1}^n  \frac{\alpha}{1 + G(m)} = \frac{\alpha^n}{Z} \prod\limits_{m=1}^n  \frac{1}{1 + G(m)},
\end{equation}
where we have introduced $\alpha = \frac{2r\epsilon^2}{\kappa\Gamma^2}$, $G(m) = F(m)/\Gamma^2$, and $Z = 1+\sum\limits_{n=1}^{\infty}\left(\prod\limits_{m=1}^n \frac{A_m}{C_{m-1}}\right)$, the normalization constant. We note that the factor $\alpha(1+G(n))^{-1}$ essentially ``propagates'' the probability distribution from $n$ to $n+1$. These results underlie the results of Fig. 2 and Fig. 3 of the main text. 

In Fig. S2, we expand upon Fig. 2 of the main text by showing the statistics as a function of pump for different coupling parameters, to give the reader of a clearer sense of the transition from thermal to coherent to Fock statistics.

\begin{figure}[h]
    \centering
    \includegraphics[width=\textwidth]{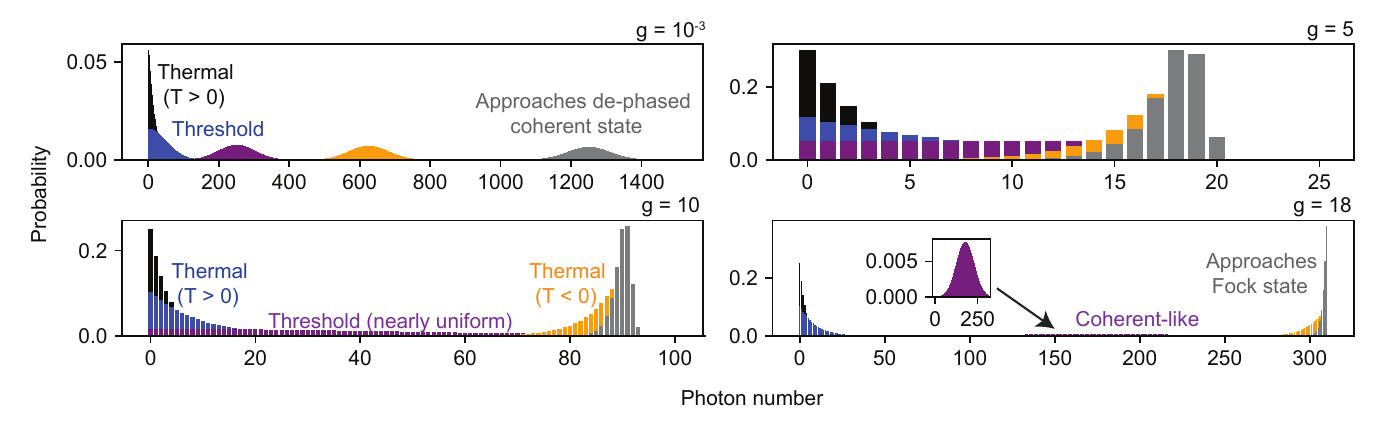}
    \caption{Evolution of photon statistics with pumping: thermal, coherent, anti-thermal, and Fock states. Photon probability distributions as a function for different pump strengths and different coupling strengths. For weak coupling, the statistics evolve from thermal to coherent with increasing pump. For the largest couplings considered, the state evolves from thermal (for low pump) to coherent (for intermediate pump) to a thermal state of \emph{negative temperature} for higher pump. As the pump increases, the negative temperature state converges effectively to a Fock state. Note that the bottom left panel overlaps with Fig. 2 of the main text. }
\end{figure}

\subsection{Summary of changes with $\lambda \neq 0$} %not sure exactly where this will go, but summarizing now

In most of the discussion, we have assumed that $\lambda = 0$. However, in many superconducting qubit systems, a nonzero $\sigma_x$ term is present due to an applied bias field which can tune the system. Our main results of nonperbative nonlinearity, as well as the Fock lasing action in these systems, are robust to the addition of this term. We summarize the main changes here. 

We assume that the generalized Rabi Hamiltonian now takes the full form
\begin{equation}
    H = \frac{1}{2}(\omega \sigma_z + \lambda \sigma_x) + \omega a^\dagger a + \tilde{g} \sigma_x (a + a^\dagger). 
\end{equation}
The spectrum is now approximately given by:
\begin{equation}
    E_{n,\sigma} = n\omega + \frac{\sigma}{2}\sqrt{(\omega D_n)^2 + \lambda^2},
\end{equation}
which corresponds to the eigenstates
\begin{align}
    \ket{n,+} &= \cos(\theta/2) D^\dagger(g) \ket{+x,n} + \sin(\theta/2)D(g)\ket{-x,n} \\
    \ket{n,-} &= \sin(\theta/2)D^\dagger(g)\ket{+x,n} - \cos(\theta/2)D(g)\ket{-x,n},
\end{align}
where the mixing angle $\theta$ is defined by $
\tan(\theta) = \omega D_n / \lambda$. 

In principles, the modifications to the analysis of the laser action should follow through new additions to the matrix elements which couple these eigenstates. Specifically, we have
\begin{align}
    \braket{n',+|(a + a^\dagger)|n,+} &= (\sqrt{n} \delta_{n',n-1} + \sqrt{n+1}\delta_{n',n+1}) - 2g\cos\theta \delta_{nn'} \\
    \braket{n',-|(a + a^\dagger)|n,-} &= (\sqrt{n} \delta_{n',n-1} + \sqrt{n+1}\delta_{n',n+1}) + 2g\cos\theta \delta_{nn'} \\
    \braket{n',-|(a + a^\dagger)|n,+} &= -2g\sin\theta \delta_{nn'}.
\end{align}
However, we see that the only new terms are only nonzero when the photon number stays the same. Thus, the only modifications to the equations of motion come from the eigen-energies. This means that the equations of motion derived previously still hold valid, but with a new nonlinearity:
\begin{equation}
    F(n) = 4n\epsilon^2  + \frac{\omega^2}{4}\left(\sqrt{D_n^2 + \lambda^2} - \sqrt{D_{n-1}^2 + \lambda^2}\right)^2.
\end{equation}

\subsection{Direct method for evolving the density matrix}

In the previous section, we treated the emitter-field interaction as if excited two-level systems were being injected into the system at rate $r$. We also treated the interaction with the emitters as sequential: as if one emitter interacts with the field at any given time, with probability coefficients averaged over the emitter's exponential decay probability. In this section, we provide an alternative treatment of the problem in which we consider the direct evolution of the density matrix in the presence of coherent emitter-field interaction, emitter pumping, emitter decay, and field leakage. This approach, besides being in principle more rigorous, and besides providing further corroboration of our results above, also allows us to consider multi-level emitter systems, such as three- and four-level systems, which are more practical from the standpoint of lasers. This method has been applied to describe conventional lasers (see \cite{scully1999quantum}), but due to its generality, can be used to describe the Fock laser discussed in this paper.

The equation of motion for the density matrix is
\begin{equation}
    \dot{\rho} = -i[H/\hbar,\rho] + \sum\limits_i \frac{\gamma_i}{2}\left(2J_i\rho J_i^{\dagger} - J_i^{\dagger}J_i\rho - \rho J_i^{\dagger}J_i \right) \equiv -i[H,\rho] + \sum\limits_i \mathcal{L}_i[\rho],
\end{equation}
where 
\begin{equation}
    H = H_{\text{Rabi}} + \sum\limits_{i=1}^N H_{\text{em,i}} + \hbar\epsilon_i(\ket{a_i}\bra{b_i} + \ket{b_i}\bra{a_i})(b+b^{\dagger}),
\end{equation}
is the Hamiltonian describing $N$ multi-level emitters (with Hamiltonian $H_{\text{em,i}}$) coupled to the electromagnetic field associated with matter strongly coupled to a single electromagnetic field mode with coupling constant $\epsilon_i$. The levels $a$ and $b$ of the $i$th emitter are coupled to the field and comprise respectively the excited and ground levels of the lasing transition. We have changed $e, g \rightarrow a, b$ as in what follows, we will introduce more levels to incorporate realistic decay channels. The terms on the right of the density matrix equation of motion are Lindblad terms with decay rates $\gamma_i$ and jump operators $J_i$. The index $i$ enumerates over the possible decay mechanisms, as well as all of the emitters. In what follows, we review several (standard) simplifications of this equation that render a readily solvable problem. 

For simplicity, we will consider the case (as before) where all emitters have the same energy levels (and thus the same $H_{\text{em,i}}$) and coupling constant $\epsilon_i = \epsilon$ (which is the average coupling dictated by the emitter distribution and cavity mode profile). Similarly, the decay constants of all atomic levels are taken as the same. These simplifications do not negate the effects reported here. Beyond these simplifications, a key simplification arises because the emitter-field coupling couples all of the emitters to a single quantum oscillator. In this case, we can consider the problem as effectively a one-emitter problem where
\begin{equation}
    H = H_{\text{Rabi}}  + \hbar\epsilon(\ket{e}\bra{g} + \ket{g}\bra{e})(a+a^{\dagger}) \equiv H_0 + V,
\end{equation}
and the $i$ in the Lindblad terms enumerates only over decay channels. We have defined for simplicity $H_0 = H_{\text{Rabi}} + H_{\text{em}}$ and $V/\hbar = \epsilon(\ket{e}\bra{g} + \ket{g}\bra{e})(a+a^{\dagger})$.

To start, we will consider decay channels for the emitter only, and not the field, and include the field decay channels at the end of the calculation. In what follows, we consider an emitter system consisting of lasing levels $a,b$, ground level $g$, and ``bath levels" $c, d$ for which $a$ and $b$ respectively decay to. The pumping from $g \rightarrow a$ occurs with rate $r$, while the $a \rightarrow c$ decay occurs with rate $\gamma_a$, the $b \rightarrow d$ decay occurs with rate $\gamma_b$, the $c \rightarrow g$ decay occurs with rate $\gamma_c$, and the $d \rightarrow g$ decay occurs with rate $\gamma_d$. Thus, the density matrix equation of motion may be written as 
\begin{equation}
    \dot{\rho} = -i[H_0/\hbar,\rho] -i[V/\hbar,\rho] + \sum\limits_{i=g,a,b,c,d} \mathcal{L}_i[\rho],
\end{equation}
where the jump operators for $g,a,b,c,d$ are respectively $\ket{a}\bra{g},\ket{c}\bra{a},\ket{d}\bra{b},\ket{g}\bra{c},\ket{g}\bra{d}$ with corresponding rates $r, \gamma_a, \gamma_b, \gamma_c, \gamma_d$.

Let us now write an equation of motion for the matrix elements of the density matrix, $\rho_{\beta n',\alpha n}$, where $\alpha, \beta$ enumerate over emitter states $g, a-d$ and the $n,n'$ enumerate over the eigenstates of the Rabi Hamiltonian (e.g., the Fock states of DSC photons). We are considering the Hamiltonian only in one spin projection, as in the previous treatment, since the spins decouple, both in the conventional Rabi model, and the generalized one (with $\lambda \neq 0$). To proceed, we will need the following matrix elements
\begin{align}
    \bra{\beta,n'}[H_0/\hbar,\rho]\ket{\alpha, n} &= (\omega_{\beta n'}-\omega_{\alpha n})\rho_{\beta n',\alpha n}\nonumber \\
    \bra{b,n'}V\rho\ket{\alpha, n} &= V_{bn',an'-1}\rho_{an'-1,\alpha n} \nonumber \\
    \bra{a,n'}V\rho\ket{\alpha, n} &= V_{an',bn'+1}\rho_{bn'+1,\alpha n} \nonumber \\
    \bra{\beta,n'}\rho V\ket{b, n} &= \rho_{\beta n',an-1}V_{an-1,bn} \nonumber \\
    \bra{\beta,n'}\rho V\ket{a, n} &= \rho_{\beta n',bn+1}V_{bn+1,an}.
\end{align}
For matrix elements of $V\rho$ and $\rho V$, we have used the structure of the matrix elements in the section "Time-evolution of the coupled system", where we showed that the effect of the coupling is to change the emitter state, and to change the number of field quanta by one. 

We also need the matrix elements of the Lindblad terms. Let us consider a generic Lindblad term of the form 
\begin{equation}
    \bra{\beta,n'}\mathcal{L}_i[\rho]\ket{\alpha, n} = \frac{\gamma_i}{2}\bra{\beta,n'}2T_{ji}\rho T^{\dagger}_{ji} - T^{\dagger}_{ji}T_{ji}\rho - \rho T^{\dagger}_{ji}T_{ji} \ket{\alpha, n},
\end{equation}
where $T_{ji} = \ket{j}\bra{i}$. $T_{ij}$ is simply $J_i$ with the final-state index $j$ included for clarity. The matrix element follows as
\begin{equation}
    \bra{\beta,n'}\mathcal{L}_i[\rho]\ket{\alpha, n} = \frac{\gamma_i}{2}(2\delta_{j\beta}\delta_{j\alpha}\rho_{in',in} - \delta_{i\alpha}\rho_{\beta n',i n} - \delta_{i\beta}\rho_{in',\alpha n}).
\end{equation}

With these matrix elements tabulated, we may write the following set of equations for the matrix elements of the density matrix:
\begin{align}
    \dot{\rho}_{an',an} &= -i\omega_{n'n}\rho_{an',an} - \gamma_a\rho_{an',an} + r\rho_{gn',gn}  - i(V_{an',bn'+1}\rho_{bn'+1,an}-\rho_{an',bn+1}V_{bn+1,an})  \nonumber \\ 
    \dot{\rho}_{bn'+1,an} &= \left[-i(\omega_{bn'+1}-\omega_{an})-\frac{\gamma_a+\gamma_b}{2} \right]\rho_{bn'+1,an} - i(V_{bn'+1,an'}\rho_{an',an}-\rho_{bn'+1,bn+1}V_{bn+1,an}) \nonumber \\
    \dot{\rho}_{an',bn+1} &= \left[-i(\omega_{an'}-\omega_{bn+1})-\frac{\gamma_a+\gamma_b}{2} \right]\rho_{an',bn+1} - i(V_{an',bn'+1}\rho_{bn'+1,bn+1}-\rho_{an',an}V_{an,bn+1}) \nonumber \nonumber \\
    \dot{\rho}_{bn'+1,bn+1} &= -i\omega_{n'+1,n+1}\rho_{bn'+1,bn+1} - \gamma_b\rho_{bn'+1,bn+1} - i(V_{bn'+1,an'}\rho_{an',bn+1}-\rho_{bn'+1,an}V_{an,bn+1})  \nonumber \\
    \dot{\rho}_{cn',cn} &=  (-i\omega_{n'n} - \gamma_c)\rho_{cn',cn} + \gamma_a \rho_{an', an}  \nonumber \\
    \dot{\rho}_{dn',dn} &= (-i\omega_{n'n} - \gamma_d)\rho_{dn',dn} + \gamma_b \rho_{bn', bn} \nonumber \\
    \dot{\rho}_{gn',gn} &=  (-i\omega_{n'n} - r)\rho_{gn',gn} +  \gamma_c\rho_{cn',cn} + \gamma_d\rho_{dn',dn}.
\end{align}
While these equations can be generally solved, we focus as in the previous treatment on the steady state dynamics. As expected from conventional lasers, the steady state density matrix is diagonal due to decoherence. Numerically, for this laser system, based on deep strong light-matter coupling, we also found that the steady-state (found by the null eigenvector of the Liouvillian ($S$ such that $\dot{\rho} = S\rho$)) is diagonal. Let us thus focus on the steady-state equations for the ``photon diagonals" ($n=n'$), which are simply 

\begin{align}
    0 &= r\rho_{gn,gn} - \gamma_a\rho_{bn,an} - i(V^*\rho_{bn+1,an}-\rho_{an,bn+1}V) \nonumber \\
    0 &= \left[i\Delta_{n+1}-\frac{\gamma_a+\gamma_b}{2} \right]\rho_{bn+1,an} - i(V\rho_{an,an}-\rho_{bn+1,bn+1}V) \nonumber \\
    0 &= \left[-i\Delta_{n+1}-\frac{\gamma_a+\gamma_b}{2} \right]\rho_{an,bn+1} - i(V^*\rho_{bn+1,bn+1}-\rho_{an,an}V^*) \nonumber \nonumber \\
    0 &= -\gamma_b\rho_{bn+1,bn+1} - i(V\rho_{an,bn+1}-\rho_{bn+1,an}V^*)  \nonumber \\
    0 &=  -\gamma_c\rho_{cn,cn} + \gamma_a \rho_{an, an}  \nonumber \\
    0 &=  -\gamma_d\rho_{dn,dn} + \gamma_b \rho_{bn, bn} \nonumber \\
    0 &=  -r\rho_{gn,gn} +  \gamma_c\rho_{cn,cn} + \gamma_d\rho_{dn,dn},
\end{align}
where we have defined $\Delta_{n+1} = \omega_{an}-\omega_{bn+1}$.

Immediately, we have $\gamma_a \rho_{an, an} = \gamma_c\rho_{cn,cn}$ and $\gamma_b \rho_{bn, bn} = \gamma_d\rho_{dn,dn}$. The equation for $\rho_{gn,gn}$ then can be written as
\begin{equation}
    r\rho_{gn} = \gamma_a\rho_{an,an} + \gamma_b\rho_{bn,bn}.
\end{equation}
For simplicity, let us take $\gamma_a = \gamma_b = \Gamma$, so that  
\begin{equation}
    r\rho_{gn} = \Gamma(\rho_{an,an} + \rho_{bn,bn}) = \Gamma(\rho_{nn}-\rho_{cn,cn}-\rho_{dn,dn}-\rho_{gn,gn}),
\end{equation}
where we have defined the photon populations $\rho_{nn} = (\text{tr}_\text{em}\rho)_{nn}$ in order to express everything in terms of these populations and arrive at a coarse-grained density matrix for the field. Let us now consider the case where $\gamma_c \gg \gamma_a$ and $\gamma_d \gg \gamma_b$. In this case, we immediately see that $\rho_{cn,cn} \approx 0$ and $\rho_{dn,dn} \approx 0$. This is to say that these levels are depleted immediately after they are populated by the lasing levels. In this case, $\rho_{gn,gn} = \frac{\Gamma}{(r+\Gamma)}\rho_{nn}$. The steady-state equations then reduce to the simple inhomogeneous equation:
\begin{equation}
\begin{bmatrix}
\Gamma & -iV & iV^* & 0\\
-iV^* & \Gamma+i\Delta_{n+1} & 0 & iV^* \\
iV & 0 & \Gamma-i\Delta_{n+1} & -iV \\
0 & iV & -iV^* & \Gamma\\
\end{bmatrix}\begin{bmatrix}
\rho_{an,an} \\
\rho_{an,bn+1} \\
\rho_{bn+1,an} \\ 
\rho_{bn+1,bn+1} \\
\end{bmatrix} = \frac{r\Gamma \rho_{nn}}{r+\Gamma}\begin{bmatrix}
1 \\
0 \\
0 \\ 
0 \\
\end{bmatrix} \equiv r_a\rho_{nn}\begin{bmatrix}
1 \\
0 \\
0 \\ 
0 \\
\end{bmatrix},
\end{equation}
whose solution yields $\rho_{an,an}, \rho_{an,bn+1}, \rho_{bn+1,an},\rho_{bn+1,bn+1}$ in terms of $\rho_{nn}$. To proceed most efficiently, we now connect these density matrix elements to the equation of motion for the reduced density matrix of the field. This equation of motion is
\begin{equation}
    \dot{\rho}_{nn} = -i(\rho_{an-1,bn}V_{bn,an-1} + V_{an,bn+1}\rho_{bn+1,an} - \rho_{bn,an-1}V_{an-1,bn} - V_{bn+1,an}\rho_{an,bn+1}). 
\end{equation}
The coherences can be found from the matrix equation above, which for brevity, we denote as $M_nP_n = r_a\rho_{nn}e_1$ so that $P_{n} = r_a\rho_{nn}(M_n^{-1}e_1)$. We may write this equation in a form similar to that of the equation for the coarse-grained density matrix of the previous treatment, i.e., as 
\begin{equation}
    0 = A_n\rho_{n-1,n-1} - A_{n+1}\rho_{n,n},
\end{equation}
where $A_n =  -ir_a((e^T_2M_{n-1}^{-1}e_1)V_{bn,an-1} - (e^T_3M_{n-1}^{-1}e_1)V_{an-1,bn})$, and we have looked at the steady-state limit. At this stage, we now add the Lindblad terms corresponding to the cavity leakage. As per the discussion in the section ``Lindblad terms", the resulting equation of motion for the field density matrix is 
\begin{equation}
    \dot{\rho}_{nn} = A_n\rho_{n-1,n-1} - (A_{n+1} + \kappa |\langle n-1|a+a^{\dagger}|n \rangle|^2)\rho_{n,n}  + \kappa |\langle n|a+a^{\dagger}|n+1 \rangle|^2\rho_{n+1,n+1}.
\end{equation}
Defining $B_n = - (A_{n+1} + \kappa |\langle n-1|a+a^{\dagger}|n \rangle|^2)$ and $C_n = \kappa |\langle n|a+a^{\dagger}|n+1 \rangle|^2$, we have as before $A_{n+1}+B_n+C_{n-1} = 0$, enabling us to immediately write
\begin{equation}
   \rho_{n+1} = \frac{A_{n+1}}{C_n}\rho_{n} \implies \rho_n = \left(\prod\limits_{m=1}^n \frac{A_m}{C_{m-1}}\right)\rho_0.
\end{equation}
with the initial $\rho_0$ is taken as one with the understanding that we must normalize the probability distribution at the end of the calculation. Upon inversion of the matrix $M_n$, we immediately find
\begin{equation}
    A_n = \frac{2r_a|V_{bn,an-1}|^2}{\Gamma^2 + 4|V_{bn,an-1}|^2 + \Delta_n^2}.
\end{equation}
Noting that $|V_{bn,an-1}|^2 = \epsilon^2 |\langle n-1|a+a^{\dagger}|n \rangle|^2$, we may write the overall equation as
\begin{equation}
    \rho_n = \frac{1}{Z}\left(\prod\limits_{m=1}^n \frac{2r_a\epsilon^2/\kappa\Gamma^2}{1+ (4\epsilon^2|\langle m-1|a+a^{\dagger}|m \rangle|^2  + \Delta_m^2)/\Gamma^2}\right) \equiv \frac{\alpha^n}{Z}\left(\prod\limits_{m=1}^n \frac{1}{1+ G(m)}\right),
\end{equation}
where $\alpha = 2r_a\epsilon^2/\kappa\gamma^2$ and $Z = 1+\sum\limits_{n=1}^{\infty}\left(\prod\limits_{m=1}^n \frac{A_m}{C_{m-1}}\right)$. Immediately, we see that if we take $|\langle n-1|a+a^{\dagger}|n \rangle|^2 = n$ and $\Delta_n^2 = \frac{1}{4}\omega^2e^{-4g^2}(L_n(4g^2)-L_{n-1}(4g^2))^2$ (assuming $\omega_0 = \omega$) that we recover the results of the previous treatment. And it may also be easily seen that this agreement persists if we take the matrix elements and splitting to be governed by the generalized Rabi model (with $\lambda \neq 0$).

\section{Numerical calculation of the Fock laser steady state}

In this section, we numerically validate the analytical developments of the previous sections. Since the analytical calculations make use of many approximations and assumptions, it is important to validate them in terms of a method which is independent of these assumptions. In what follows, we will use a method inspired by the observation that the equation of motion for the laser density matrix effectively describes the interaction of a single gain atom with a cavity, even when the gain medium is composed of many atoms. This is because the atoms only couple to each other through the cavity field, as noted in \cite{scully1999quantum}. Thus, it follows that laser steady states can be understood through the steady state of the Liouvillian operator describing a damped oscillator coupled to a gain atom. By taking the partial trace of the null eigenvector of the Liouvillian, one finds the steady state probability distribution of DSC photons. Applied to conventional lasers, one correctly finds the transition from thermal to coherent state statistics above the laser threshold. 

The Hamiltonian part of the Liouvillian is simply the Hamiltonian of Eq. (2). Here, we also consider two different possible interaction terms: $a+a^{\dagger}$ or $b+b^{\dagger}$. The steady-states are also insensitive to this. In these calculations, we include a reservoir to describe pumping of the gain medium, as well as its $T_1$- and $T_2$-relaxation (here, $T_2 = 2T_1$). We also include a reservoir to describe the decay of the DSC photon. We take as the jump operator $J^{(+)}$, where $J = a+a^{\dagger}$ or $b+b^{\dagger}$ and the $+$ superscript means ``positive-frequency'', meaning that we project out all negative frequency components. The steady-state is insensitive to whether we use $a$ or $b$ (it changes only slightly), indicating the relative unimportance of the dipole-dipole interaction from the standpoint of the photon probabilities. The overall Liouvillian is then the sum of the Hamiltonian (commutator) part and three dissipators: $\mathcal{D}[\sigma_{\text{em}}^{(-)}], \mathcal{D}[\sigma_{\text{em}}^{(+)}]$ and $\mathcal{D}[J^{(+)}]$ with respective rates $r$, $\Gamma$, and $\kappa$, to describe gain pumping, gain decoherence, and DSC photon decay. Note that $\mathcal{D}[O]\rho \equiv O^{\dagger}O\rho + \rho O^{\dagger}O - 2O\rho O^{\dagger}$. 

In Fig. S3, we show the probability distribution of DSC photons resulting from one of these steady-state calculations. Because we have largely neglected spin in our analytical discussions, we plot the ``unpolarized'' photon probability distribution, defined such that $P(n) = P(n,-1)+P(n,1)$. As can be seen, above threshold, the state has very low number fluctuations, in this case, $\delta n$ = 1, yielding a state very close to a Fock state. Below threshold, a thermal state is found. Around threshold, the quasi-uniform state of Fig. 3 of the main text is found. 

These results are insensitive to the presence of a $\lambda$-term, as shown in Fig. S4. The presence of a $\lambda$ term, all else equal, slightly increases the photon noise. This is because the presence of a $\lambda$ softens the anharmonicity (which can be understood from the term $\sqrt{\lambda^2 + \omega^2 e^{-4g^2}L^2_n(4g^2)}$ in Eq. (60)).

The results are also insensitive to the exact form of interaction and dissipator (provided that the dissipator doesn't create spurious excitations). In Fig. S5, we show the steady-state, computed using interaction terms based on $a$ or $b$, as well as dissipators based on $a$ or $b$. 

\begin{figure}[h]
    \centering
    \includegraphics[width=0.35\textwidth]{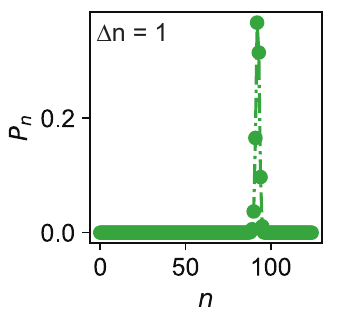}
    \caption{Steady-state of the Fock laser, calculated numerically, by finding the steady-state of the Liouvillian operator. Plot shows the unpolarized probability distribution for $\lambda = 0$. For $\epsilon = 10^{-5}\omega$, $\kappa = 10^{-8}\omega$, and $r = 10\Gamma$ (such that the population inversion of the gain is about 90\%, the resulting state is nearly a Fock state of 100 DSC photons, with a residual uncertainty of 1. This state has noise 99\% below the shot noise level. Moreover, this calculation shows that the Hamiltonian of Eq. (3), coupled to damping, supports Fock states as its steady state, from first principles. }
\end{figure}

\begin{figure}[h]
    \centering
    \includegraphics[width=0.7\textwidth]{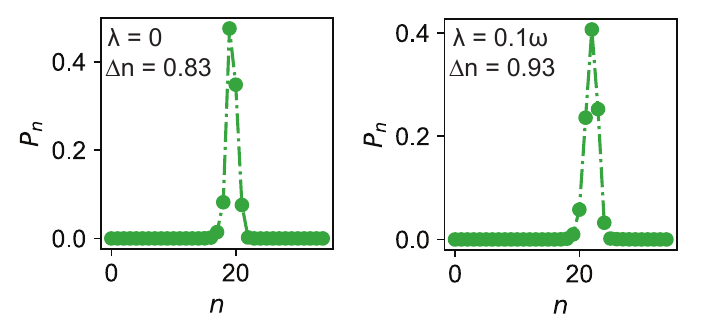}
    \caption{Steady-state of the Liouvillian of the Fock laser with and without the $\lambda$-term of Eq. (1) of the main text. Here, $\lambda = 0.1\omega$ is sufficient to keep the two spin ladders from interchanging, and is not found to alter the steady-state appreciably. $\epsilon, \Gamma, \kappa$ are the same as in the above figure. }
\end{figure}

\begin{figure}[h]
    \centering
    \includegraphics[width=0.7\textwidth]{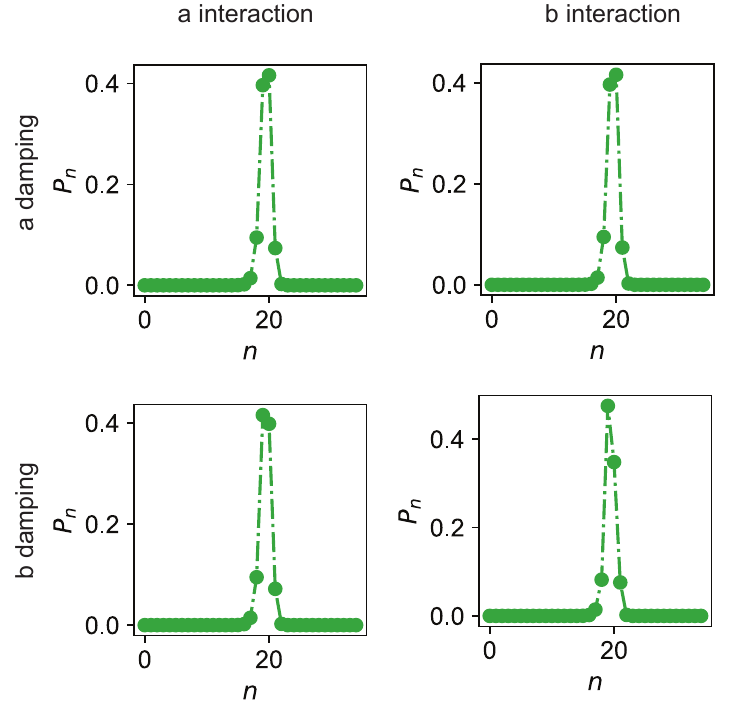}
    \caption{Steady-state of the Liouvillian of the Fock laser with different choices for the interaction term and dissipator, showing robustness to the exact magnitude of the dipole-dipole interaction between the emitter and the qubit. }
\end{figure}

\clearpage

\section{Coherent pumping of sharp DSC nonlinearity in the presence of loss}

In the main text we showed how coherently pumped DSC bosons can show a striking effect of $N$-photon blockade due to sharp anharmonicity, generating highly nonclassical states of light. For simplicity of discussion, the main text results did not include the effects of loss. We report on the effects of loss here, which were studied through solving the master equation associated with coherently pumped DSC bosons in the presence of damping. 

To do so, we consider the equations of motion associated with a coherently pumped nonlinear oscillator. We assume that the driven nonlinear oscillator has a Hamiltonian
\begin{equation}
    H / \hbar = \sum_n \omega_n \ket{n}\bra{n} + \eta \cos(\omega t) (a + a^\dagger),
\end{equation}
where $\eta$ is the drive strength, and $\omega$ is the drive frequency. Additionally, $\omega_n$ is the spectrum of the anharmonic oscillator, which for the DSC system is computed using the methods described in earlier sections of the paper. In a frame which rotates with the frequency of the drive, and after making the rotating wave approximation, the pumped system has the time independent Hamiltonian
\begin{equation}
    \tilde{H} = \sum_n \delta_n \ket{n}\bra{n} + \frac{\eta}{2}(a + a^\dagger).
\end{equation}
Here, $\delta_n$ is a detuning from the pump frequency defined as $\omega_n = n\omega + \delta_n.$ In this sense, $\delta_n$ encodes the anharmonicity of the spectrum with respect to the coherent driving frequency. The time evolution in this rotating picture is given by the density matrix equation $\partial_t \tilde{\rho} = i[\tilde{\rho}, \tilde{H}] + \mathcal{L}[\tilde{\rho}] =\equiv L[\tilde{\rho}].$ Here, $L[\tilde{\rho}]$ is the Liouvillian operator acting on the density matrix. Additionally, $\mathcal{L}[\tilde{\rho}]$ is a standard Lindblad term which performs linear damping on the $a$ operator with a rate $\kappa$. From this, we can compute time evolution of the states, or solve the steady state condition $L[\tilde{\rho}] = 0.$ 

Although the Hamiltonian is not exactly the same as the driven Hamiltonian presented in the main text in the absence of loss, we find that the models behave nearly identically in the limit of low loss, indicating that the simplified model presented here provides a good description of the key physics. The main advantage of this approach is that in the rotating frame, the Hamiltonian becomes time independent, enabling numerical solutions of the steady state solution that do not require time evolution of the density matrix. 

\begin{figure}[h]
    \centering
    \includegraphics[width=0.9\textwidth]{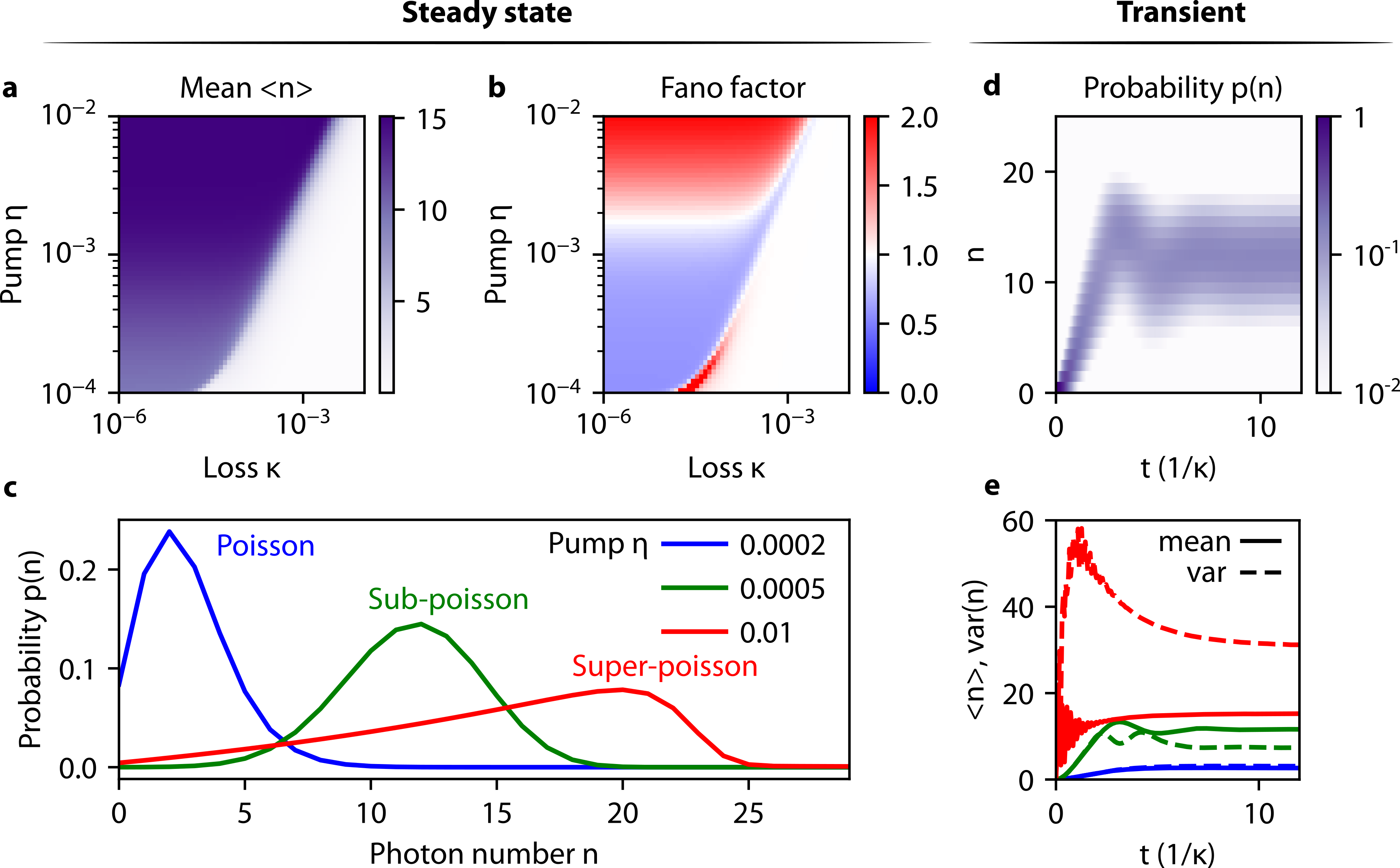}
    \caption{\textbf{Deep strong coupling bosons pumped by a coherent driving field.} Coupling strength $g=5$ is used across all panels. (a) Steady state mean excitation number $\langle n \rangle$ for various pump strength $\eta$ and cavity damping factor $\kappa$. (b) Steady state Fano factor $f = \langle (\Delta n)^2 \rangle / \langle n \rangle$ for the same values of $\eta$ and $\kappa$. For low pump strengths, $f=1$, indicating typical coherent state pumping. For higher pump strengths, the steady state cavity statistics exhibit both sub-poissonian ($f < 1$) and super-poissonian ($f > 1$) behavior. (c) Steady state probability distributions $p(n)$ for three chosen pump values $\eta = \{2\times 10^{-4}, 5\times 10^{-4}, 10^{-2} \}$ for a fixed damping rate $\kappa = 10^{-4}$. The three pump rates respectively lead to poissonian, sub-poissonian, and super-poissonian steady state statistics due to their varying interactions with the nonlinear cavity. (d) Transient evolution of the probability distribution $p(n)$ of the pumped system starting from the ground state at $t=0$, with parameters $\eta = 5 \times 10^{-4}$, $\kappa = 10^{-4}$. (e) Transitent evolution of the excitation number mean and variance for the same pump parameters shown in (c). At any given time, the fano factor can be extracted by noting how the variance compares to the mean. Additionally, nonlinear oscillations between the mean and variance are seen as system reaches steady state.}
    \label{fig:dsc_pumping_loss}
\end{figure}

Figure \ref{fig:dsc_pumping_loss} shows both steady state and transient behaviors for DSC bosons pumped with coherent drive strength $\eta$ and loss rate $\kappa$. Evidence of the blockade effect can be seen in Fig. \ref{fig:dsc_pumping_loss}a, where increased pumping results in a minimal increase in the mean occupation number. This occurs because states beyond the blockade number of $N= g^2 = 25$ cannot be occupied. The higher the loss $\kappa$, the larger the pump strength required to populate the states. Panel b shows how the steady state Fano factors behave as a function of the same pump and loss values. We see that when the loss is too high for a given pump, decoherence from dissipation dominates the steady state, and a coherent state with $F=1$ is recovered. However, when the loss is sufficiently low, both super-Poissonian and sub-Poissonian states can be created. Interestingly, we find that higher pumps result in super-Poissonian behavior, as more of the state is ``reflected'' back toward lower photon numbers. Representative examples of Poissonian, sub-Poissonian, and super-Poissonian behavior are shown in Fig. \ref{fig:dsc_pumping_loss}c.

We can also see how these states are created transiently. Fig. \ref{fig:dsc_pumping_loss}d shows how the representative sub-Poissonian (green) state is created when pumped from the ground state, in terms of its probability distribution. In the early stages of time evolution, the mean photon number grows quadratically in time, as it would for an ordinary harmonic oscillator. This occurs since at early times, the pumped field only occupies the parts of the DSC boson spectrum which are essentially perfectly harmonic. However, once the tail of the probability distribution approaches the blockade number $N$, nonlinear oscillations begin to occur, as the state is reflected from the boundary. Eventually, these oscillations damp out, and a steady distribution is reached, corresponding to the green curve in Fig. \ref{fig:dsc_pumping_loss}c. We observe that in the presence of losses, it appears that the minimum attainable Fano factor is around 0.5. 

Fig. \ref{fig:dsc_pumping_loss}e shows how the mean and variance evolve as a function of time for the states shown in Fig. \ref{fig:dsc_pumping_loss}c. We see that the lowest pump level generates a coherent state in a manner nearly identical to a linear cavity. The subpoissonian state is generated with influence from the blockade, as described above, and a few cycles of nonlinear oscillation are seen in the mean and variance as the steady state is approached. Finally, the strongest coherent pump results in a super-Poissonian distribution, which is reached after rapid nonlinear oscillations. We note that even though the mean photon number is only marginally higher than that of the sub-Poissonian example state, the variance is higher by more than a factor of 3.

We finally comment on the key differences between these results and the lossless calculations shows in the main text. One key difference is that the lossless calculations in the main text exhibit a periodic ``revival" behavior as the probability distribution oscillates between the ground state and the blockade number. Instead of these huge oscillations, the damped system exhibits oscillations as it finally settles into a steady state. Relatedly, the presence of loss eliminates the presence of fringes in the probability distribution as the state hits the blockade number and squeezes. In this sense, the loss has the effect of moderating the extent of quantum features which can be realized in the state. This is entirely consistent with the well-known fact that dissipation acts as a form of decoherence which degrades quantum states. Nevertheless, even in the presence of loss, this coherently pumped system exhibits clear features of $N$-photon blockade, and can be used to generate DSC boson quantum states with sub-Poissonian character. 

\clearpage

\clearpage

\bibliographystyle{unsrt}
\bibliography{Fock_laser.bib}